\documentclass{aa}
\usepackage{graphicx}
\usepackage{txfonts}
\usepackage{subfigure}
\usepackage{multirow}
\usepackage{natbib}
\usepackage{array}
\usepackage{soul}
\usepackage[normalem]{ulem}

\usepackage{lscape}
\begin{document}
   \title{Chemical Enrichment RGS cluster sample (CHEERS): \\Constraints on turbulence}


   \author{Ciro Pinto
          \inst{1}
          \and
          Jeremy~S. Sanders \inst{2}
          \and
          Norbert Werner \inst{3,4}
          \and
          Jelle de Plaa \inst{5}
          \and
          Andrew~C. Fabian \inst{1}
          \and
          Yu-Ying Zhang \inst{6}
          \and\\
          Jelle~S. Kaastra\inst{5}
          \and
          Alexis Finoguenov \inst{7}
          \and
          Jussi Ahoranta \inst{7}
          }

   \institute{Institute of Astronomy, Madingley Road, CB3 0HA Cambridge, United Kingdom, \email{cpinto@ast.cam.ac.uk}.
         \and
             Max-Planck-Institut fur extraterrestrische Physik, Giessenbachstrasse 1, D-85748 Garching, Germany
         \and
             Kavli Institute for Particle Astrophysics and Cosmology, Stanford University, 452 Lomita Mall, Stanford, CA 94305-4085, USA
         \and
             Department of Physics, Stanford University, 382 Via Pueblo Mall, Stanford, CA 94305-4060, USA
         \and
             SRON Netherlands Institute for Space Research, Sorbonnelaan 2, 3584 CA Utrecht, The Netherlands.
         \and
             Argelander-Institut f\"ur Astronomie, Universit\"at Bonn, Auf dem H\"ugel 71, 53121 Bonn, Germany. 
         \and
             Department of Physics, University of Helsinki, FI-00014 Helsinki, Finland
             }
    \date{Received 5 November 2014 / Accepted 24 December 2014}

 
  \abstract
   {Feedback from active galactic nuclei, galactic mergers, and sloshing are thought  
   to give rise to turbulence, which may prevent cooling in clusters.}
   {We aim to measure the turbulence in clusters of galaxies and
   compare the measurements to some of their structural and evolutionary properties.}
   {It is possible to measure the turbulence of the hot gas in clusters by estimating the velocity 
    widths of their X-ray emission lines. The Reflection Grating Spectrometers 
    aboard {XMM-\textit{Newton}} are currently the only instruments provided with 
    sufficient effective area and spectral resolution in this energy domain.
    We benefited from excellent 1.6\,Ms new data provided by the Chemical Enrichment RGS cluster sample (CHEERS) project.}
   {The new observations improve the quality of the archival data and allow us to place constraints 
    for some clusters, which were not accessible in previous work. 
  {One-half of the sample shows upper limits on turbulence less than 500\,km\,s$^{-1}$.
    For several sources, our data are consistent with
    relatively strong turbulence with upper limits on the velocity widths that are
    larger than 1000\,km\,s$^{-1}$. The NGC\,507 group of galaxies shows transonic velocities,
    which are most likely associated with the merging  
    phenomena and bulk motions occurring in this object. 
    Where both low- and high-ionization emission lines
    have good enough statistics,
    we find larger upper limits for the hot gas, which is partly due
    to the different spatial extents of the hot and cool gas phases. 
    Our upper limits are larger than the Mach numbers required to balance cooling,
    suggesting that dissipation of turbulence may prevent cooling,
    although other heating processes could be dominant.}
    {The systematics associated with the spatial profile of the source continuum 
    make this technique very challenging, though still powerful, for current instruments.}
    In a forthcoming paper we will use the resonant-scattering technique to place lower-limits 
    on the velocity broadening and provide further insights on turbulence.
    The ASTRO-H and Athena missions will revolutionize the velocity estimates  
    and discriminate between different spatial regions and temperature phases.}
   {}


   \maketitle
%

\section{Introduction}
\label{sec:introduction}
Clusters of galaxies are the most massive, individual, bound objects in the Universe.
In their gravitational potential well, the gas, called the intra-cluster medium (ICM),
is heated to temperatures of $10^{7-8}$\,K and, therefore, strongly emits at X-ray energies.
The ICM is commonly thought to be in hydrostatic equilibrium, but there are several
factors that may affect the dynamical state of the gas. The feedback from active galactic nuclei (AGN) creates bubbles 
that may drive turbulence up to about 500\,km\,s$^{-1}$ 
(see, e.g., \citealt{Bruggen2005} and \citealt{Fabian2005}). 
Sloshing of gas within the gravitational potential may produce similar velocities,
while galactic mergers can give rise to even higher velocities of about 1000\,km\,s$^{-1}$ 
(see, e.g., \citealt{Lau2009}, \citealt{Ascasibar2006}).

{The AGN feedback is thought to offset radiative losses and 
to suppress cooling in isolated giant elliptical galaxies 
and in larger systems up to the richest galaxy clusters (see, e.g., \citealt{McNamara2007}
and \citealt{Fabian2012}).
Simulations and observations have confirmed that AGN feedback 
may prevent cooling through the production of turbulence 
(see, e.g., \citealt{Ruszkowski2004}, \citealt{Zhuravleva2014}, and \citealt{Gaspari2014}). 
Other work suggests that turbulent mixing may also be an important mechanism 
through which AGN heat cluster cores (see, e.g., \citealt{Banerjee2014}).}

\vspace{-0.05cm}

It is possible to measure velocity broadening on the order of few hundreds km\,s$^{-1}$ directly in the X-ray emission lines
produced by the hot ICM. The Reflection Grating Spectrometers (RGS, \citealt{denherder2001}) aboard 
XMM-\textit{Newton} are currently the only X-ray instruments, which have enough collecting area and spectral resolution
to enable this measurement.
However, the spatial extent of clusters complicates the process due to the slitless
nature of the RGS.
\citet{Sanders2010} made the first measurement of cluster velocity broadening
using the luminous cluster \object{A\,1835} at redshift 0.25. Due to the limited spatial extent 
of its bright core, an upper limit of 274\,km\,s$^{-1}$ was obtained.
\citet{Sanders2011} then constrained turbulent velocities for a large sample
of 62 sources observed with XMM-\textit{Newton}/RGS, which included clusters, groups, and elliptical galaxies.
Half of them show velocity broadening below 700\,km\,s$^{-1}$. Recently, 
\citet{Sanders2013} used continuum-subtracted emission line surface brightness
profiles to account for the spatial broadening. This technique is affected by systematic errors of
up to 150\,km\,s$^{-1}$.

\citet{Werner2009} and \citet{dePlaa2012} measured turbulent velocities through the ratio of the 
\ion{Fe}{xvii} emission lines at 15 and 17\,{\AA}. When the velocity broadening is low, the gas is optically thick 
in the 15\,{\AA} line due to resonant scattering, while the 17\,{\AA} lines remain optically thin. The comparison
of observed with simulated line ratios for different Mach numbers constrains the level of turbulence.
This method is very efficient for cool core clusters rich in \ion{Fe}{xvii} emission lines, but it is partly
limited by the systematic uncertainty ($\sim$20\%) in the line ratio for an optically thin plasma.

In this work, we measure the velocity broadening
for the 44 sources of the CHEmical Enrichment RGS cluster Sample (CHEERS),
which is connected to a Very Large Program accepted for XMM-\textit{Newton} AO-12. 
We model the line spatial broadening using CCD images.
This method has systematics due to the spatial profile 
of the continuum, which may overestimate the line spatial broadening,
but it is still a useful technique to measure the level of velocity broadening
when deep, high-spatial resolution maps are lacking. 
We also test an alternative method, 
which uses a variable spatial-broadening.
The paper is organized as follows. 
In Sect.\,\ref{sec:cheers}, we give a brief description of the CHEERS project. 
In Sect.\,\ref{sec:data}, we present the data reduction. 
Our method is described in Sect.\,\ref{sec:spectral_modeling}. 
We discuss the results in Sect.\,\ref{sec:discussion} and 
give our conclusions in Sect.\,\ref{sec:conclusion}.
Further useful material is reported in Appendix\,\ref{sec:appendix}
to speed up the paper reading.

\vspace{-0.35cm}

\section{The CHEERS project}
\label{sec:cheers}

The current catalog includes 44 nearby, bright clusters, groups of galaxies, and elliptical galaxies
with a value of a $\gtrsim5\sigma$ detection for the \ion{O}{viii} 1s--2p line at 19\,{\AA}
and with a well-represented variety of strong, weak, and non cool-core objects. 
This catalog also contains 19 new observations of 1.6\,Ms in total, which are taken during AO-12, 
PI: J. de Plaa (see Table~\ref{table:log}). 
More detail on the sample choice is provided by another paper 
(de Plaa et al., in preparation). 
Among the several goals of this large project, we mention the following ones:
\vspace{-0.2cm}
\begin{itemize}
 \item To understand the ICM metal enrichment by different SN types, \
       (see, e.g., Mernier et al. accepted) 
 \item to study substructures, asymmetries and multiphaseness,
 \item to study heating and cooling in cluster cores,
 \item to measure turbulence (this paper),
 \item to improve the cross-calibration between X-ray satellites.
\end{itemize}

\vspace{-0.5cm}

\section{Data}
\label{sec:data}

The data used in this paper are listed in Table~\ref{table:log}. 
In boldface, we show the new observations
taken during AO-12. A few archival exposures have not been used, 
since they were too short. 

The XMM-\textit{Newton} satellite is equipped with two types of X-ray detectors: 
The CCD-type European Photon Imaging Cameras (EPIC) and the Reflection Grating Spectrometers (RGS). 
The European photon imaging cameras are
MOS\,1, MOS\,2, and pn (\citealt{Struder2001} and \citealt{Turner2001}). 
The RGS camera consists of two similar detectors, which have both high effective area and 
spectral resolution between 6 and 38\,{\AA} \citep{denherder2001}.
The MOS cameras are aligned with the RGS detectors and have 
higher spatial resolution than the pn camera. 
We have used MOS\,1 for imaging and RGS for spectral analysis.


\subsection{RGS and MOS 1 data reduction}

The data were reduced with the XMM-\textit{Newton} Science Analysis System (SAS) v13.5.0. 
We processed the RGS data 
with the SAS task \textit{rgsproc} and the MOS\,1 data with \textit{emproc} 
{to produce event files, spectra, and response matrices for RGS and MOS data}.

To correct for contamination from soft-proton flares, we {used the SAS task \textit{evselect}
to extract} light curves for MOS\,1 in the 10--12 keV
energy band, while we used the data from CCD number 9 for RGS 
where hardly any emission from each source is expected. We
binned the light curves in 100\,s intervals. A Poissonian distribution was fitted to the
count-rate histogram, and all time bins outside the $2\sigma$ level were rejected.
We built the good time intervals (GTI) files with the accepted time events for the MOS and RGS files 
{through the SAS task \textit{tabgtigen} and 
reprocessed the data again with \textit{rgsproc} and \textit{emproc}}. 
The RGS\,1 total clean exposure times are quoted in Table\,\ref{table:log}.

\begin{table*}
\caption{XMM-\textit{Newton}/RGS observations used in this paper.}  
\vspace{-0.25cm}
\label{table:log}      
\renewcommand{\arraystretch}{1.1}
 \small\addtolength{\tabcolsep}{+2pt}
 
\scalebox{1}{%
\begin{tabular}{c c c c c c c }     
\hline\hline            
Source                               &  ID $^{(a)}$          & Total clean time (ks) $^{(b)}$ & $kT$ (keV) $^{(c)}$  & $z\,^{(c)}$  & $N_{\rm H}$ ($10^{24}\,{\rm m}^{-2}$) $^{(d)}$\\  
\hline   
\multirow{1}{*}{\object{2A0335+096}} & 0109870101/0201 0147800201               &  120.5      &    3.0      &  0.0349    & 30.7  \\
\multirow{1}{*}{\object{A 85}}       &  \textbf{0723802101/2201}                &  195.8      &    6.1      &  0.0556    & 3.10  \\
\multirow{1}{*}{\object{A 133}}      &  0144310101 \textbf{0723801301/2001}     &  168.1      &    3.8      &  0.0569    & 1.67  \\
\multirow{1}{*}{\object{A 189}}      &  0109860101                              &   34.7      &    1.3      &  0.0320    & 3.38  \\
\multirow{1}{*}{\object{A 262}}      &  0109980101/0601 0504780101/0201         &  172.6      &    2.2      &  0.0161    & 7.15  \\
\multirow{1}{*}{\object{A 496}}      &  0135120201/0801 0506260301/0401         &  141.2      &    4.1      &  0.0328    & 6.00  \\
\multirow{1}{*}{\object{A 1795}}     & 0097820101                               &   37.8      &    6.0      &  0.0616    & 1.24  \\
\multirow{1}{*}{\object{A 1991}}     & 0145020101                               &   41.6      &    2.7      &  0.0586    & 2.72  \\
\multirow{1}{*}{\object{A 2029}}     & 0111270201 0551780201/0301/0401/0501     &  155.0      &    8.7      &  0.0767    & 3.70  \\
\multirow{1}{*}{\object{A 2052}}     & 0109920101 0401520301/0501/0601/0801     &  104.3      &    3.0      &  0.0348    & 3.03  \\
                                     & 0401520901/1101/1201/1301/1601/1701      &             &             &            &   \\
\multirow{1}{*}{\object{A 2199}}     & 0008030201/0301/0601 \textbf{0723801101/1201} &  129.7 &    4.1      &  0.0302    & 0.909  \\
\multirow{1}{*}{\object{A 2597}}     & 0108460201 0147330101 \textbf{0723801601/1701} & 163.9 &    3.6      &  0.0852    & 2.75  \\
\multirow{1}{*}{\object{A 2626}}     & 0083150201 0148310101                    &   56.4      &    3.1      &  0.0573    & 4.59  \\
\multirow{1}{*}{\object{A 3112}}     & 0105660101 0603050101/0201               &  173.2      &    4.7      &  0.0750    & 1.38  \\
\multirow{1}{*}{\object{A 3526}}     &  0046340101 0406200101                   &  152.8      &    3.7      &  0.0103    & 12.2  \\
\multirow{1}{*}{\object{A 3581}}     &  0205990101 0504780301/0401              &  123.8      &    1.8      &  0.0214    & 5.32  \\
\multirow{1}{*}{\object{A 4038}}     & 0204460101 \textbf{0723800801}           &   82.7      &    3.2      &  0.0283    & 1.62  \\
\multirow{1}{*}{\object{A 4059}}     & 0109950101/0201 \textbf{0723800901/1001} &  208.2      &    4.1      &  0.0460    & 1.26  \\
\multirow{1}{*}{\object{AS 1101}}    & 0147800101 0123900101                    &  131.2      &    3.0      &  0.0580    & 1.17  \\
\multirow{1}{*}{\object{AWM 7}}      & 0135950301 0605540101                    &  158.7      &    3.3      &  0.0172    & 11.9  \\
\multirow{1}{*}{\object{EXO 0422}}   & 0300210401                               &   41.1      &    3.0      &  0.0390    & 12.4  \\
\multirow{1}{*}{\object{Fornax}}     & 0012830101 0400620101                    &  123.9      &    1.2      &  0.0046    & 1.56  \\
\multirow{1}{*}{\object{HCG 62}}     & 0112270701 0504780501 0504780601         &  164.6      &    1.1      &  0.0140    & 3.76  \\
\multirow{1}{*}{\object{Hydra-A}}    & 0109980301 0504260101                    &  110.4      &    3.8      &  0.0538    & 5.53  \\
\multirow{1}{*}{\object{M 49}}       & 0200130101                               &   81.4      &    1.0      &  0.0044    & 1.63  \\
\multirow{1}{*}{\object{M 86}}       & 0108260201                               &   63.5      &    0.7      &  -0.0009   & 2.97  \\
\multirow{1}{*}{\object{M 87} (Virgo)} & 0114120101 0200920101                    &  129.0      &    1.7      &  0.0042    & 2.11  \\
\multirow{1}{*}{\object{M 89}}       & 0141570101                               &   29.1      &    0.6      &  0.0009    & 2.96  \\
\multirow{1}{*}{\object{MKW 3s}}     & 0109930101 \textbf{0723801501}           &  145.6      &    3.5      &  0.0450    & 3.00  \\
\multirow{1}{*}{\object{MKW 4}}      & 0093060101 \textbf{0723800601/0701}      &  110.3      &    1.7      &  0.0200    & 1.88  \\
\multirow{1}{*}{\object{NGC 507}}    &  \textbf{0723800301}                     &   94.5      &    1.3      &  0.0165    & 6.38  \\
\multirow{1}{*}{\object{NGC 1316}}   & 0302780101 0502070201                    &  165.9      &    0.6      &  0.0059    & 2.56  \\
\multirow{1}{*}{\object{NGC 1404}}   & 0304940101                               &   29.2      &    0.6      &  0.0065    & 1.57  \\
\multirow{1}{*}{\object{NGC 1550}}   & 0152150101 \textbf{0723800401/0501}      &  173.4      &    1.4      &  0.0123    & 16.2  \\
\multirow{1}{*}{\object{NGC 3411}}   & 0146510301                               &   27.1      &    0.8      &  0.0152    & 4.55  \\
\multirow{1}{*}{\object{NGC 4261}}   & 0056340101 0502120101                    &  134.9      &    0.7      &  0.0073    & 1.86  \\
\multirow{1}{*}{\object{NGC 4325}}   & 0108860101                               &   21.5      &    1.0      &  0.0259    & 2.54  \\
\multirow{1}{*}{\object{NGC 4374}}   & 0673310101                               &   91.5      &    0.6      &  0.0034    & 3.38  \\
\multirow{1}{*}{\object{NGC 4636}}   & 0111190101/0201/0501/0701                &  102.5      &    0.8      &  0.0037    & 2.07  \\
\multirow{1}{*}{\object{NGC 4649}}   & 0021540201 0502160101                    &  129.8      &    0.8      &  0.0037    & 2.23  \\
\multirow{1}{*}{\object{NGC 5044}}   & 0037950101 0584680101                    &  127.1      &    1.1      &  0.0090    & 6.24  \\
\multirow{1}{*}{\object{NGC 5813}}   & 0302460101 0554680201/0301/0401          &  146.8      &    0.5      &  0.0064    & 6.24  \\
\multirow{1}{*}{\object{NGC 5846}}   & 0021540101/0501 \textbf{0723800101/0201} &  194.9      &    0.8      &  0.0061    & 5.12  \\
\multirow{1}{*}{\object{Perseus}}    & 0085110101/0201 0305780101               &  162.8      &    6.8      &  0.0183    & 20.7  \\
\hline                
\end{tabular}}

$^{(a)}$ Exposure ID number. $^{(b)}$ RGS net exposure time. 
$^{(c)}$ Redshifts and temperatures are adapted from \cite{Chen2007} and \cite{Snowden2008}. 
$^{(d)}$ Hydrogen column density (see http://www.swift.ac.uk/analysis/nhtot/).
New observations from our proposal are shown in boldface.\\
      \vspace{-0.5cm}
\end{table*}


\subsection{RGS spectra extraction}
\label{sec:rgs_regions}

\begin{figure}
  \begin{center}
      \subfigure{ 
      \includegraphics[bb=15 15 515 348, width=7.5cm]{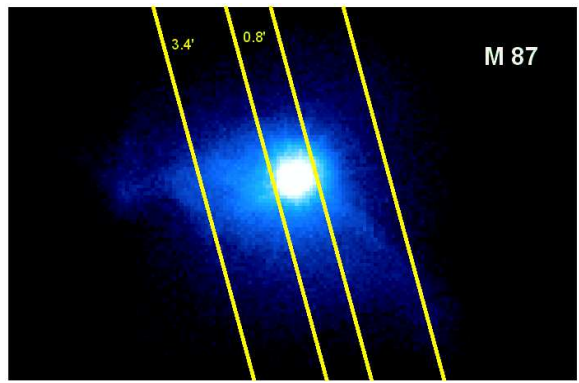}}
      \vspace{-0.25cm}
      \caption{RGS extraction regions and MOS\,1 stacked image of M\,87.}
          \label{fig:rgs_regions}
  \end{center}
      \vspace{-0.5cm}
\end{figure}

We extracted the RGS source spectra in two alternative regions centered on the emission peak:
a broader 3.4' region, which includes most of the RGS field of view and a narrower 0.8' region that provides
the cluster cores but with high statistics. {This was done by launching \textit{rgsproc} twice
by setting the \textit{xpsfincl} mask to include 99\% and 90\% of point-source events 
inside the spatial source extraction mask, respectively.} We have used the model background spectrum 
created by the standard RGS \textit{rgsproc} pipeline, which is a template background file,
based on the count rate in CCD\,9. 
The RGS spectral extraction regions and the MOS\,1 image of M\,87 are shown in Fig.~\ref{fig:rgs_regions}. 
The spectra were converted to SPEX\footnote{www.sron.nl/spex} format through the SPEX task \textit{trafo}.
{During the spectral conversion, we chose the option of \textit{sectors} in the task \textit{trafo}
to create as many sectors as the different exposures of each source. This permits us to simultaneously
fit the multiple RGS spectra of each source by choosing which parameters to either couple or unbind 
in the spectral models of different observations.}

 \vspace{-0.5cm}

\subsection{MOS 1 spatial broadening profiles}
\label{sec:spatial_profile}

The RGS spectrometers are slitless, and, therefore,
the spectra are broadened because of the spatial extent of the source in the dispersion direction. 
The effect of this spatial broadening is described by the following wavelength shift
\begin{equation}
\Delta\lambda = \frac{0.138}{m} \, \Delta\theta \, {\mbox{\AA}},
\end{equation}
where $m$ is the spectral order and $\theta$ is the offset angle of the source in arcmin
(see the XMM-\textit{Newton} Users Handbook).

\begin{figure}
  \begin{center}
      \subfigure{ 
      \includegraphics[bb=66 66 536 707, width=6cm, angle=90]{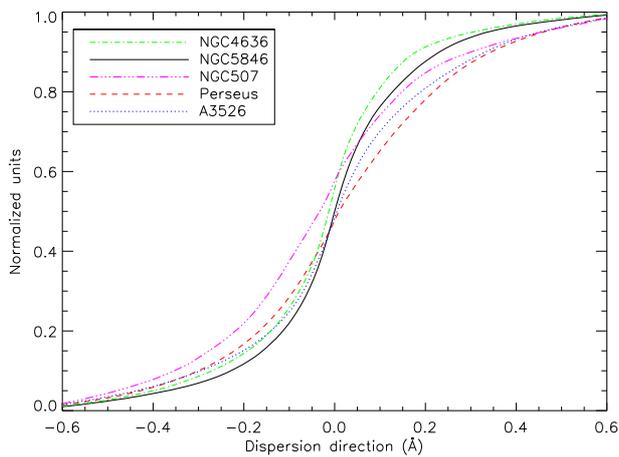}}
      \vspace{-0.3cm}
      \caption{MOS\,1 average 7--25\,{\AA} cumulative spatial profiles.}
          \label{fig:profiles}
  \end{center}
      \vspace{-0.4cm}
\end{figure}

The MOS\,1 DET\,Y direction is parallel to the RGS\,1 dispersion direction 
and can be used to correct for the spatial broadening. 
{With \textit{evselect}}, we extracted MOS\,1 images for each exposure in the 0.5--1.8\,keV
(7--25\,{\AA}) energy band and their surface brightness profiles in the dispersion
direction.
We account for spatial broadening using the \textit{lpro} multiplicative model in SPEX,
which convolves the RGS response with our model of the spatial extent of the source. 
We show some cumulative profiles of spatial broadening in Fig.~\ref{fig:profiles}.
We have also produced stacked Fe-L band (10--14\,{\AA}) images for each source. 
The central 10' region contains most of the cluster emission 
(see Fig.~\ref{fig:mos1}). 

\vspace{-0.4cm}


\section{Spectral modeling}
\label{sec:spectral_modeling}

Our analysis focuses on the $7-28$ {\AA} ($0.44-1.77$ keV) first and second order RGS spectra.
We model the spectra with SPEX
version 2.03.03. 
We scale elemental abundances to the proto-Solar values 
of \citet{Lodders09}, which are the default in SPEX. 
We adopt C-statistics and $1\,\sigma$ errors throughout the paper,
unless otherwise stated, and the updated ionization balance calculations of \citet{Bryans2009}.


Clusters of galaxies are not isothermal, and most of them have both hot and cool gas phases 
(see e.g. \citealt{Frank2013}). Therefore, we have used a two-temperature thermal plasma model
of collisional ionization emission (CIE). This model is able to fit all the spectra in our database. 
The \textit{cie} model in SPEX calculates the spectrum of a plasma 
in collisional ionization equilibrium. The basis for this
model is given by the mekal model, but several updates have been included (see the SPEX manual).
Free parameters in the fits are the emission measure $Y=n_{\rm e}\,n_{\rm H}\,dV$, the temperature $T$, 
the abundances (N, O, Ne, Mg, and Fe), and the turbulent broadening $v$ of the two \textit{cie} models.

We bind the parameter $v$ and the abundances of two \textit{cie} components with each other
and assume that the gas phases have the same turbulence and abundances. 
This decreases the degree of degeneracy. 
{This assumption is certainly not true, but some clusters just need one CIE component
and the spectra of several clusters do not have good enough statistics in both high- and low-ionization emission lines,
which prohibits constraining the velocities and the abundances for both hot and cool phases.
We attempt to constrain the turbulence in the different phases in Sect.\,\ref{sec:temperature}.}

The \textit{cie} models are multiplied by a \textit{redshift} model 
and a model for Galactic absorption, which is provided by the \textit{hot} model in SPEX 
with $T=0.5$\,eV and $N_{\rm H}^{TOT}$, as estimated through the tool of \citet{Willingale2013}. 
This tool includes the
contribution to absorption from both atomic and molecular hydrogen. The redshifts and
column densities that have been adopted are shown in Table~\ref{table:log}.
To correct for spatial broadening, we have multiplied the spectral model 
by the \textit{lpro} component that receives as input the surface brightness profile
extracted in the MOS\,1 images (see Sect.\,\ref{sec:spatial_profile} and Fig.\,\ref{fig:profiles}).

We do not explicitly model the cosmic X-ray background in the RGS spectra 
because any diffuse emission feature 
would be smeared out into a broad continuum-like component. 

For a few sources, including the Perseus and \object{Virgo} (M\,87) clusters, 
we have added a further power-law emission component
of slope $\sim2$ to take the emission from the central AGN into account.
This component is not convolved with the spatial profile because 
it is produced by a central point-like source.

To avoid the systematic effects due to the stacking of multiple observations
with different pointing,
we have simultaneously fitted the individual spectra
of each source extracted in the two regions defined in Sect.~\ref{sec:rgs_regions} 
and shown in Fig.\,\ref{fig:rgs_regions}. 
The plasma model is coupled between the observations. 
The only uncoupled parameters are the emission measures 
of the two collisional-ionized gas components. 
For each observation we adopt the spatial profile extracted in the MOS\,1 image taken during that exposure.
For those exposures, during which the MOS\,1 detector had a closed filter, 
we have adopted an exposure-weighted average profile as given by the other available observations, 
but the $\delta\lambda$ parameter in the \textit{lpro} component is left free. 
This factor allows us to shift the model lines by the same amount (in {\AA}) for each specific spectrum
and strongly decreases the systematic effects.
The $\delta\lambda$ parameter is always free in our fits to account for any redshift variation,
which would otherwise affect the line modeling (see e.g. \citealt{Sanders2011}).

The simultaneous modeling of multiple observations has been done through the use
of the \textit{sectors} option in SPEX (see also Sect.\,\ref{sec:rgs_regions}).
The RGS\,1 and 2 spectra of the same observation
have exactly the same model and provide a single sector, while RGS spectra of other observations
contribute additional sectors and have the \textit{cie} normalizations uncoupled.

\subsection{{Results using a fixed spatial broadening}}
\label{sec:results}

We have successfully applied this multi-temperature model to both the 3.4' and 0.8' RGS spectra.
We show the spectral modeling for the 3.4' region of the 44 sources 
in Figs.~\ref{fig:rgs_fits}, \ref{fig:rgs_fits2}, and \ref{fig:rgs_fits3} in Appendix\,\ref{sec:appendix}. 
We display the first-order stacked spectra to underline the high quality of these observations
and to show the goodness of the modeling.

For some sources like Fornax, M\,49, M\,86, NGC\,4636, and NGC\,5813, 
the 15 and 17\,{\AA} \ion{Fe}{xvii} emission lines are not well fitted. 
Precisely, the model underestimates the line peaks and overestimates the broadening.
This may be due to the different spatial distribution of the gas responsible for the cool \ion{Fe}{xvii} 
emission lines and for the one producing most of the high-ionization Fe-L and \ion{O}{viii} lines. 
The cool gas is indeed to be found predominantly in the center of the clusters 
showing a profile more peaked than that one of the hotter gas. 
The estimated spatial profiles depend on the emission of the hotter gas due to its higher emission measure,
and, therefore, they overestimate the spatial broadening of the 15--17\,{\AA} lines.
It is hard to extract a spatial profile for these lines because MOS\,1 has a limited spectral resolution,
and the images extracted in such a short band lack the necessary statistics 
(see e.g. \citealt{Sanders2013}). {In Sect.\,\ref{sec:temperature}, we attempt to constrain the turbulence
for lines of different ionization states.} The 15\,{\AA}\,/\,17\,{\AA} line ratio is also affected 
by resonant scattering, which would require a different approach. 
We refer to a forthcoming paper on the analysis of the resonant scattering in the CHEERS sources.

We skip the discussion of the abundances and the supernova yields 
because these will be treated by other papers 
of this series (de\,Plaa et al. in preparation and Mernier et al. submitted).

In Fig.\,\ref{fig:turbulence2} (\textit{\textit{left panel}}), we show the upper limits on the velocity broadening obtained 
with the simultaneous fits of the 0.8' 7--28\,{\AA} RGS spectra.
We obtain upper limits for most clusters, while 
NGC\,507 shows high kinematics. {More detail on our results for the 3.4' and 0.8' regions 
and their comparison are reported in Table\,\ref{table:velocity_results} and Fig.\,\ref{fig:velocities_comparison} (\textit{\textit{left panel}}).
The 3.4' limits are more affected by the source continuum,
as clearly seen for M\,87, AWM\,7, and A\,4038,
which makes them less reliable.

\begin{figure*}
  \begin{center}
      \subfigure{ 
      \includegraphics[bb=110 77 535 723, width=9cm]{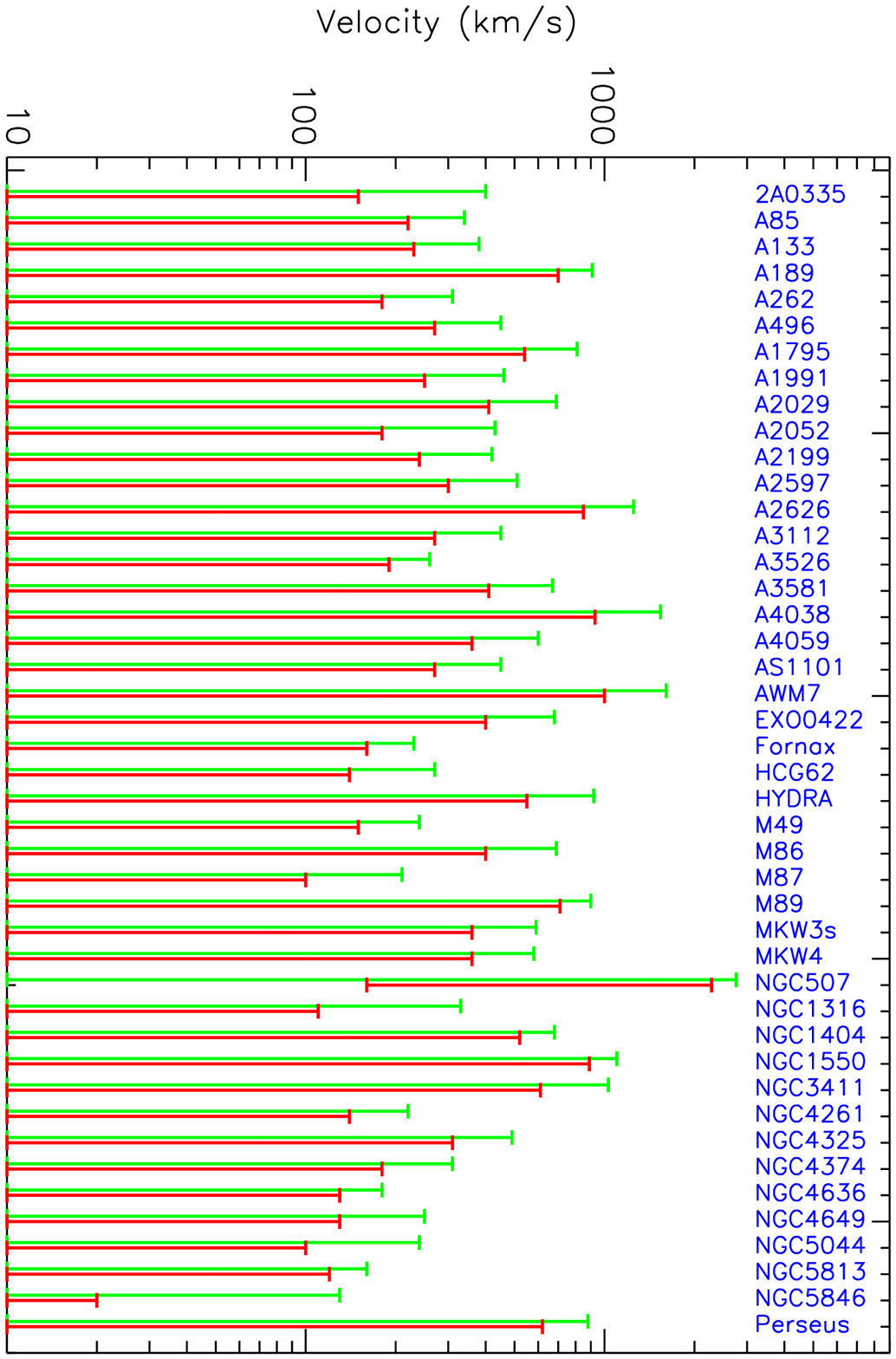} \hspace{0cm}  
      \includegraphics[bb=110 77 535 723, width=9cm]{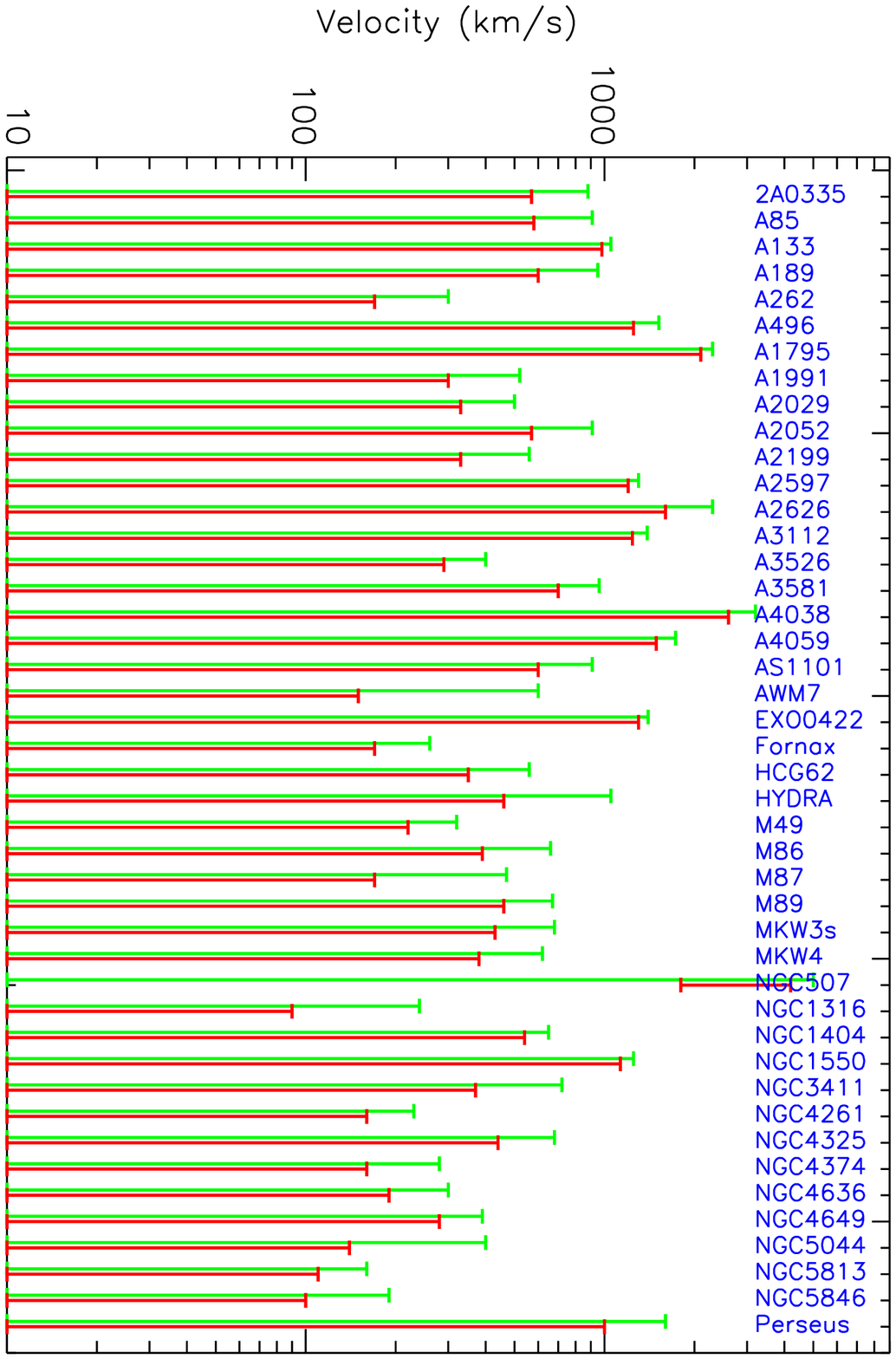} }
      \vspace{-0.2cm}
      \caption{\textit{Left panel}: Velocity 68\% (red) and 90\% (green) limits for the 0.8' region
               with the spatial broadening determined with MOS\,1 images (see Sect.\,\ref{sec:results}). 
               \textit{Right panel}: Velocity limits obtained using the best-fit spatial broadening
               (see Sect\,\ref{sec:results_combined}).}
          \label{fig:turbulence2}
  \end{center}
      \vspace{-0.6cm}
\end{figure*}

\subsection{{Results using the best-fit spatial broadening}}
\label{sec:results_combined}

It is known that the spatial profile of the source continuum may be
broader than the spatial distribution of the lines. The MOS\,1 
images are strongly affected by the profile of the source continuum
and, therefore, may overestimate the spatial line broadening
and underestimate the residual velocity broadening. 
For instance, NGC\,1316 and NGC\,5846 
show $1\sigma$ limits of 20\,km\,s$^{-1}$, which are not realistic (see Table\,\ref{table:velocity_results}).

To obtain more conservative limits, we have simultaneously modeled
the spatial and the velocity broadening. This was done by fitting the RGS 0.8' spectra
with a free \textit{s} parameter in the \textit{lpro} component.
This factor simply scales the width of the spatial broadening 
by a factor free to vary (see the SPEX manual).
The free \textit{s} parameter increases the degeneracy in the model 
but provides conservative upper limits on the residual velocity broadening,
which is measured with the $v$ parameter of the \textit{cie} component.
The new limits on the velocities are plotted in Fig.\,\ref{fig:turbulence2} (\textit{right panel})
and quoted in the last two columns of Table\,\ref{table:velocity_results}.
In Fig.\,\ref{fig:velocities_comparison} (\textit{right panel}), we compare the velocity upper limits
estimated with the standard method (MOS\,1 spatial profile with $s\equiv1$ in the \textit{lpro} component)
with this new approach using a free $s$ parameter. They generally agree,
but the new upper limits on the hotter Abell clusters
are systematically larger by an average factor $\sim2$. 
This confirms that some of the previous velocity limits were 
underestimated due to the broader spatial profiles.
Therefore, we believe the new upper limits to be the most conservative.
Interestingly, the conservative velocity limits of the hot clusters 
are generally higher than the cool galaxy groups with the exception of NGC\,507,
which is expected since the sound speed scales as a power
of the temperature (see Sect.\,\ref{sec:turbulence}).

      \vspace{-0.4cm}

\subsection{Further tests}
\label{sec:tests}

To estimate the contribution of the spatial broadening to the line widths,
we have temporarily removed the convolution of the spectral model 
for the spatial profile and re-fitted the data. 
In these fits the $v$ parameter of the \textit{cie} component 
accounts for any contribution to the line broadening. 
The total (spatial + velocity) widths are also quoted in Table\,\ref{table:velocity_results}.

We have also tested the continuum-subtracted line surface brightness profiles 
introduced by \cite{Sanders2013}. This new method consists of subtracting the surface brightness profiles
of two regions that are clearly line-dominated (core) and continuum-dominated (outskirts).
It can be applied only to those objects with a narrow core 
where it is possible to distinguish between line-rich and line-poor regions.
We have locally fitted the \ion{O}{viii} 19.0\,{\AA} emission line of 
A\,2597, A\,3112, Hydra-A, Fornax (\object{NGC\,1399}), and NGC\,4636 
and we have found a general agreement with the results of \cite{Sanders2013}.
However, our MOS\,1 images have much lower spatial resolution than the \textit{Chandra} maps
used by them, which increases the uncertainties that are present in this method. 
A thorough, extensive, analysis would require deep \textit{Chandra} maps 
that are not yet available.



\section{Discussion}
\label{sec:discussion}

In this work we have analyzed the data of 44 clusters, groups of galaxies, and elliptical galaxies
included in the CHEERS project, a Very Large Program that was accepted 
for XMM-\textit{Newton} AO-12 (see Sect.\,\ref{sec:cheers}) 
together with complementary archival data.

We have measured upper limits of velocity broadening for these objects
with a method similar to the previous one used by \citet{Bulbul2012} and \citet{Sanders2013}. 
This consists of fitting high-quality grating spectra
by removing the spatial broadening through surface brightness profiles of the sources as provided
by CCD imaging detectors. 
These profiles are unfortunately affected by the source continuum
and tend to overestimate the line spatial broadening with a consequent
shrinking of the residual velocity broadening. 
\citet{Sanders2013} addressed this point in their Sect.\,2.2 on A\,3112
where they decreased these systematic effects by using \textit{Chandra} continuum-subtracted
line spatial profiles. We have tested this method by using the MOS\,1 observations
that were taken simultaneously with the RGS spectra (see Fig.\,\ref{fig:mos1}). 
XMM-\textit{Newton} CCDs have a spatial resolution lower than \textit{Chandra} CCDs, 
which increase the systematic effects in the creation of continuum-subtracted maps.
Deep \textit{Chandra} observations, enabling an accurate  
subtraction of different energy bands, are missing for most sources.
We have therefore tried to use the MOS\,1 integral maps
and to fit the contribution of the spatial broadening
as an alternative method.


\subsection{Temperature dependence of the upper limits}
\label{sec:temperature}

{So far we adopted the same velocity broadening for all the emission lines.
For most sources it is possible to measure the velocity broadening of the 
\ion{O}{viii} and \ion{Fe}{xx-to-xxiv} emission lines, which are mainly produced by hot gas.
Only a few sources have high-statistics \ion{Fe}{xvii} lines produced by cool ($T<1$\,keV) gas.
Six objects exhibit both strong low- and high-ionization lines 
and allow to fit the velocity broadening of the two \textit{cie} components, separately,
in the full-band spectral fits.
17 sources allow to measure 90\% upper limits on turbulence
for the \ion{O}{viii}, \ion{Fe}{xvii}, and \ion{Fe}{xx} lines, 
by fitting the $18.0-23.0$\,{\AA}, $14.0-18.0$\,{\AA}, and $10.0-14.3$\,{\AA} 
rest-frame wavelength ranges, respectively. 
For each local fit we adopt an isothermal model and correct for spatial broadening
by using additional surface-brightness profiles calculated through MOS\,1 images
extracted in the same rest-frame wavelength ranges using the same method
shown in Sect.\,\ref{sec:spatial_profile}. These profiles are still affected by the 
continuum but provide a better description of the spatial broadening in each line.
In Fig.\,\ref{fig:turb_vs_band} we compare the \ion{O}{viii} velocity limits
with those measured for the \ion{Fe}{xvii} and \ion{Fe}{xx} line systems.
The high-ionization Fe lines clearly show higher upper limits,
which is confirmed by the results of the full-band fits: 
the hotter (T1) CIE component allows for higher values of velocity broadening.
The hotter gas is distributed over a larger extent than that of the cold (T2) gas 
and has larger spatial broadening,
which affects the T1-T2 results shown in this plot. The \ion{O}{viii}, 
\ion{Fe}{xvii}, and \ion{Fe}{xx} lines were fitted by subtracting the spatial broadening 
extracted exactly in their energy band, which should partly correct this systematic effect,
but it is difficult to estimate the systematic uncertainties
due to the low spatial (and spectral) resolution of the CCD data.
On some extent, the hotter phase may still have larger turbulence. 
For clarity, we also tabulate these line-band fits in Table\,\ref{table:physical_properties}.
We also note that the the velocity limits of low- and high-ionization iron lines
fall at opposite sides of the Fe--\ion{O}{viii} 1:1 line, which means that the metallicity 
distribution in the sources should not affect our broad-band, multi-ion, limits 
shown in Fig.\,\ref{fig:turbulence2}.}

\begin{figure}
      \subfigure{ 
      \includegraphics[bb=58 54 540 730, width=6.5cm, angle=+90]{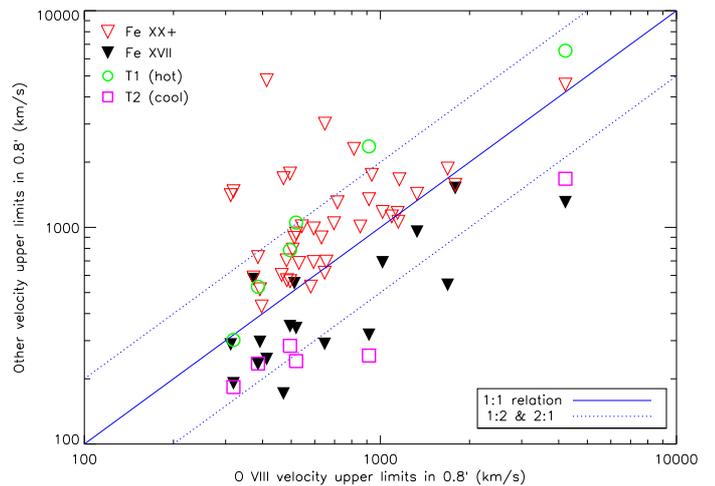}}
      \caption{{90\% upper limits on velocity broadening obtained in the 0.8' region 
               for the \ion{O}{viii} lines compared with those measured for high-ionization \ion{Fe}{xx} 
               (open red triangles) and for the low-ionization \ion{Fe}{xvii} (filled black triangles) line systems.
               For six sources we could also measure the limits for the hot (open green circles) 
               and the cool (open magenta boxes) CIE components {(see Sect.\,\ref{sec:temperature})}.
                \label{fig:turb_vs_band}}}
\end{figure}


\subsection{Turbulence}
\label{sec:turbulence}

In Fig.\,\ref{fig:turbulence2} we show the velocity broadening 
of the RGS spectra extracted in the 0.8' core region. 
We find upper limits to the velocity broadening with the possible exception of NGC\,507. 
They generally range between 200 and 600\,km\,s$^{-1}$. 
For several objects like A\,85, A\,133, M\,49, and most NGC elliptical
we found velocity levels below 500\,km\,s$^{-1}$, which would
suggest low turbulence.
The broader 3.4' region is more affected by spatial broadening 
as shown by the higher upper limits, but non-detection, of
A\,4038, AWM\,7, and M\,87 (see Table\,\ref{table:velocity_results}
and Fig.\,\ref{fig:velocities_comparison}). 
For these sources it is difficult to constrain the velocity broadening 
because their large extent smears out the emission lines.

To understand how much energy can be stored in turbulence, we 
compare our upper limits with the sound speeds and the temperatures of 
dominant \textit{cie} component in these objects.
The sound speed is given by $c_S = \sqrt{\gamma \, k \, T / \mu m_{\rm p}}$,
where $\gamma$ is the adiabatic index, which is 5/3 for ideal monoatomic gas,
$T$ is the RGS temperature, $\mu=0.6$ is the mean particle mass, and $m_p$ is proton mass.
The ratio between turbulent and thermal energy is 
$\varepsilon_{\rm turb}/\varepsilon_{\rm therm}=\gamma/2\,M^2$,
where $M=v_{\rm turb}/c_S$ is the Mach number (see also \citealt{Werner2009}).
In Fig.\,\ref{fig:velocities_comparison} we compare our $2\sigma$ upper limits 
on the velocities in the central 0.8' region
with the sound speed and some fractions of turbulent energy.
We also show the more conservative velocity upper limits
that were measured with a variable spatial broadening. 
At least for half the sample, our 90\% upper limits are below the sound speed in the system.
In about ten objects the turbulence contains less than the 40\%
of the thermal energy. This is similar to the previous results
of \cite{Sanders2013}. Apparently, the hotter objects allow for higher velocities.

{We note that the spectral extraction region had a fixed angle, and, therefore,
the actual physical scale -- where we estimated the velocity broadening -- depends on the source distance.
In Fig.\,\ref{fig:turb_vs_temp_scaled} (\textit{left panel}), we show the Mach numbers
for the 90\% conservative upper limits as a function of the temperature, and we compare
the average upper limits on Mach number calculated within different ranges of physical scales.
There is no significant trend with the temperature,
but the average upper limit on the Mach number is lower for narrower physical scales.
Assuming a Kolmogorov spectrum for the turbulence in these objects, 
the root-mean-square velocity scale depends on the 1/3rd power of the physical length.
Therefore, we scaled the upper limits by ${(sc/sc_{\rm min})}^{1/3}$, 
where $sc_{\rm min}$ is the 
minimum physical scale per arcsec $\sim0.07$\,kpc/1" of NGC\,4636,
the nearest object in our sample.
In other words, we divided our upper limits by the relative physical scale per arcsec 
relative to NGC\,4636, which is equivalent to normalizing by the ratio
between the size of the spectral extraction region of each cluster
and that one of NGC\,4636.
The scaled upper limits on the Mach numbers are tabulated 
in Table\,\ref{table:physical_properties} and plotted
in Fig.\,\ref{fig:turb_vs_temp_scaled} (\textit{right panel}).
They are randomly distributed around $Ma\sim0.8$ and 
do not depend any more on the physical scale. 
We coded the point-size and the colors with the values of $r_{500}$ and $K_0$ taken from 
the literature. The $r_{500}$ is the radius within which the mean over-density 
of the cluster is 500 times the critical density at the cluster redshift,
and $K_0$ is the value of the central entropy in the same cluster.
All the adopted values and their references are reported in Table\,\ref{table:physical_properties}.
We do not find any significant relation between the upper limits on the Mach number and 
these physical properties, possibly due to the limited sample.}

{To understand whether dissipation of turbulence may prevent cooling
in our sample, we computed the Mach number that is required to balance 
the heating and cooling, according to the following equation:}
\begin{equation}\label{eq:mach}
Ma_{REQ} \approx 0.15 \left( \frac{n_e}{10^{-2} \, {\rm cm}^{-3}} \right)^{1/3} 
                   \, \left( \frac{c_s}{10^3 \, {\rm km\,s}^{-1}} \right)^{-1} 
                   \, \left( \frac{l}{10 \, {\rm kpc}} \right)^{1/3}
\end{equation}
{where $n_e$ is the density at the cavity location, $c_s$ the sound speed
that we have estimated through the RGS temperature, 
and $l$ the characteristic eddy size, which we take as the
average cavity size (see \citealt{Zhuravleva2014}).
The Mach numbers required to balance cooling are tabulated 
in Table\,\ref{table:physical_properties}.
Most cavity sizes were taken from \citet{Panagoulia2014b}.
For clusters with multiple cavities, we used an average size.
For the 19 sources outside of their sample, we used their $r-T$ relation 
to determine the cavity size. Most densities were taken from the ACCEPT
catalog.
}

{In Fig.\,\ref{fig:Mach_vs_temp_scaled}, we compare the ratios between 
the conservative upper limits of the scaled Mach numbers assuming Kolmogorov turbulence,
and those that are required to balance cooling with the RGS temperatures.
For most sources, our upper limits are larger than the balanced Mach numbers,
which means that dissipation of turbulence can provide enough heat
to prevent the cooling of the gas in the cores.}

{It is difficult to know which is the main mechanism that produces turbulence
in these objects. Our scaled upper limits are mostly below 500\,km\,s$^{-1}$,
which can be produced by bubbles inflated by past AGN activity (see, e.g., \citealt{Bruggen2005}).
For some objects, our upper limits are consistent with velocities up to 1000\,km\,s$^{-1}$, 
which would correspond to Mach numbers larger than one.
For NGC\,507, we detect transonic motions presumably due to merging
(see, e.g., \citealt{Ascasibar2006}).
In a forthcoming paper, we will analyze the resonant scattering of the \ion{Fe}{xvii} lines
exhibited by half of our sample to place lower limits on turbulent broadening 
and provide more insights on its origin and its role in preventing cooling.}

\begin{figure}
      \subfigure{ 
      \includegraphics[bb=65 85 525 686, width=6.8cm, angle=+90]{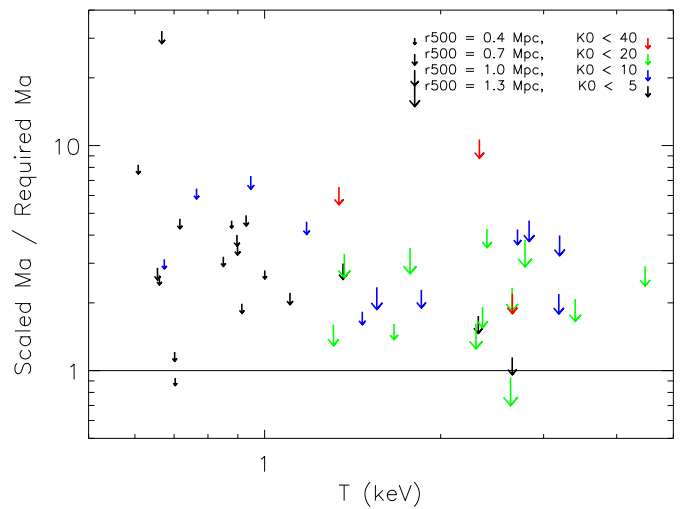}}
      \vspace{-0.5cm}
      \caption{{Ratios between the 90\% conservative upper limits on the Mach number (velocity / sound speed)
               that are scaled by the 1/3rd power of the spatial scalel assuming Kolmogorov turbulence
               (see {Sect.\,\ref{sec:turbulence}}), and the Mach number, which is required
               to make a heating--cooling balance (see Eq.\,\ref{eq:mach}).
               The point size provides the $r_{500}$, and the color is coded according to the central entropy,
               $K_0$, in units of keV cm$^{2}$.
                \label{fig:Mach_vs_temp_scaled}}}
\end{figure}


\subsection{Comparison with previous results}
\label{sec:comparison}

Our velocity limits broadly agree with the previous results obtained by \citet{Sanders2013} 
using a similar method and by other authors, who use the measurements
of resonant scattering (\citealt{Werner2009} and \citealt{dePlaa2012}).
In particular, our limits for M\,49 (also known as \object{NGC\,4472}), 
NGC\,4636, and NGC\,5813 
agree with the $100$\,km\,s$^{-1}$ upper limit obtained by \citet{Werner2009}.
We also found upper limits of a few $100$s\,km\,s$^{-1}$ for A\,3112, which is similar
to the results of \citet{Bulbul2012}.
However, we measured higher limits with a variable spatial broadening
that agree with continuum-subtracted profiles method of \citet{Sanders2013}.

Recently, \citet{Zhuravleva2014} used the surface brightness fluctuations 
in the \textit{Chandra} images of the Perseus and Virgo clusters to derive turbulent 
velocities in the range 70--210\,km\,s$^{-1}$ for Perseus and 43--140\,km\,s$^{-1}$ for Virgo,
{where the smaller values refer to the central 1.5' region.}
{Our upper limits in the cores of the clusters are consistent with their values,
especially when normalized by the physical scale factor $1.5'/0.4'$.}
They show that these turbulent motions should dissipate enough energy to 
offset the cooling of the central ICM in these clusters.
{For ten objects, the scaled Mach number can be transonic, and a major fraction of energy
can be stored in turbulence, which could significantly heat the gas through dissipation 
(see, e.g., \citealt{Ruszkowski2004}).
Recently, \citet{Gaspari2014} noted that even if the turbulence 
in the hot gas is subsonic, it may be transonic in the cooler gas phases.
\citet{Zhuravleva2014} reported that  dissipation of turbulence may balance cooling
even under subsonic regime.
Our upper limits on Mach number are larger than the values necessary to balance cooling
and are consistent with this scenario.
However, it is possible that other processes are dominant, 
such as turbulent mixing (see, e.g., \citealt{Banerjee2014}).}

The NGC\,507 group exhibits velocities larger than $1000$\,km\,s$^{-1}$ in both the 0.8' and 3.4' regions,
{corresponding to a scaled Mach number $Ma=4.2\pm1.7$ (1$\sigma$)}. 
The 15\,{\AA} \ion{Fe}{xvii} line is stronger than the one at 17\,{\AA}, which
would suggest low resonant scattering (see Fig.\,\ref{fig:rgs_fits3})
and, therefore, high kinematics in the galaxy group.
This object is known to have a disturbed shape and to host radio lobes presumably
in a transonic expansion/inflation (\citealt{Kraft2004}). However, our high values
suggest the presence of bulk motions. 
{In Fig.\,\ref{fig:NGC507}, we show the velocities of the galaxies in the NGC\,507 group
as taken from \cite{Zhang2011}. They are not necessary
linked to that of the ICM, but there are high kinematics and hints 
of infalling clumps, which indicate a substructure extended toward the observer.
In this group, the galaxy velocities generally double those observed in NGC\,4636,
where we measure lower velocity broadening (see also the different line widths in Fig.\,\ref{fig:rgs_fits3}).}

\begin{figure}
  \begin{center}
      \subfigure{ 
      \includegraphics[bb=50 115 535 590, width=7.5cm, angle=-90]{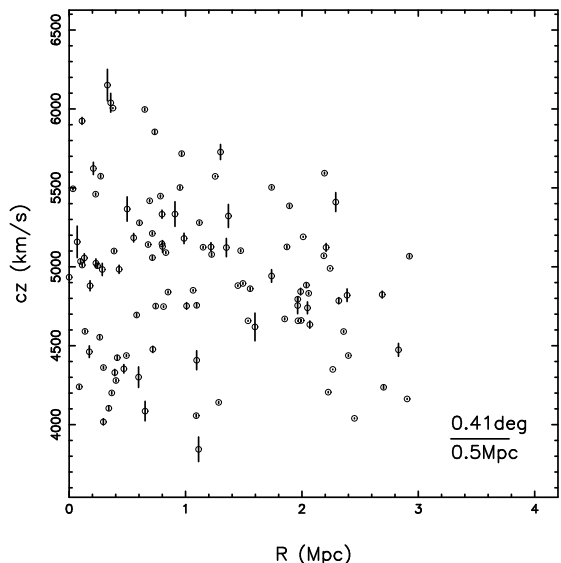}}
  \vspace{-0.4cm}
      \caption{Line-of-sight velocity versus projected distance from the central cD galaxy 
               for the member galaxies of NGC\,507 group.
               Optical spectroscopic redshifts are taken from \cite{Zhang2011}.}
          \label{fig:NGC507}
  \end{center}
\end{figure}

\subsection{Toward ASTRO-H}
\label{sec:simulations}

The RGS gratings aboard XMM-\textit{Newton} are currently the only instruments that can
measure $100$s\,km\,s$^{-1}$ velocities in X-ray spectra of extended sources like clusters of galaxies. 
However, they are slitless spectrometers and, therefore, affected by spatial broadening.
We have partly solved this issue by using line surface brightness profiles, 
but there are still systematic uncertainties larger than 100\,km\,s$^{-1}$.
Our models provide an important workbench once the new ASTRO-H
X-ray satellite (\citealt{Takahashi2010}) is launched. The spectra, as provided by its microcalorimeter (SXS),
do not suffer from spatial broadening as for the RGS and will revolutionize the method. 
Moreover, its constant spectral resolution
in terms of energy increases the sensitivity at high energies, which allows us 
to use higher-ionization lines up to 6-7\,keV (Fe-K line complex)  
necessary to constrain the turbulence in hotter gas phases. 
The position of the lines unveil evidence of bulk motions.

In Fig.\,\ref{fig:simulations} (\textit{\textit{left panel}}), we compare the effective area of the ASTRO-H SXS
with that of the first order RGS 1 and 2. ASTRO-H provides clearly better results than the sum of RGS\,1 and 2 below 14\,{\AA} 
(above 1\,keV). The RGS has still a better spectral resolution than the SXS 
in the wavelength range that includes the \ion{Fe}{xvii} lines of the cool gas,
but the absence of spurious line-broadening in the SXS makes it a great alternative tool.
We have simulated a 100\,ks exposure with the ASTRO-H SXS for four interesting objects
in our catalog: Perseus (500\,km\,s$^{-1}$), NGC\,5846 (10\,km\,s$^{-1}$), NGC\,4636 (100\,km\,s$^{-1}$), 
and NGC\,507 (1000\,km\,s$^{-1}$, see Fig.\,\ref{fig:simulations} \textit{right panel}). 
We have used the model fitted for the full (-1.7',+1.7') RGS spectra as a template, which are shown
in Fig.\,\ref{fig:rgs_fits3}, because this extraction region is 
comparable to the 3.05'\,$\times$\,3.05' field-of-view of the microcalorimeter.
The spatial broadening was excluded from the model. 
The simulated SXS spectra are characterized by a richness of resolved emission lines, which provides
velocity measurements with an accuracy of 50\,km\,s$^{-1}$ or better.
The line widths clearly increase throughout NGC\,5846, NGC\,4636, and NGC\,507.
The hotter gas present in the Perseus cluster produces strong higher-ionization lines
above 1\,keV, which constrain the turbulence in different (Fe-L and Fe-K) gas phases. 

We also note that the 1' spatial resolution of ASTRO-H provides,
for the first time the means for a spatially-resolved high-resolution spectral analysis
and the measurements of turbulence in different regions of the clusters.
The ATHENA X-ray observatory that is to be launched by the late 2020s will further 
revolutionize our measurements due to its combined high spectral (2.5\,eV) and spatial ($<5$'') resolution.

\section{Conclusion}
\label{sec:conclusion}

We have presented a set of upper limits and measurements of the velocity widths 
for the soft X-ray emitting gas of a sample of clusters, groups of galaxies, 
and elliptical galaxies included in the CHEERS project. 
We have subtracted the instrumental spatial broadening 
through the use of surface brightness profiles
extracted in the MOS\,1 images. 

For most sources, we obtain upper limits ranging within 200-600\,km\,s$^{-1}$, 
{where the turbulence may originate in AGN feedback or sloshing of the ICM.
However, for some sources, such as NGC\,507, we find upper limits of 1000\,km\,s$^{-1}$ 
or larger, suggesting other origins, such as mergers and bulk motions.
The measurements depend on the angular scale and the temperature.
For a small sample producing strong high- and low-ionization
lines, we measured significantly broader upper limits for the hot gas phase,
which may be partly due to its larger spatial extent as compared
to the cool phase.
When we normalize the Mach numbers for the physical scale, assuming 
Kolmogorov turbulence, we constrain upper limits ranging within
$0.3<Ma<1.5$. These values are above the Mach numbers necessary to balance
cooling, which means that the dissipation of turbulence
may be the dominant mechanism to heat the gas and quench cooling flows.
However, it is possible that additional processes are heating the ICM.}
In a forthcoming paper, we will use the resonant-scattering technique 
to place some lower limits on the velocities in half of our sample,
which are characterized by strong \ion{Fe}{xvii} emission lines, and 
make an extensive use of higher-resolution \textit{Chandra} maps.
This will provide alternative measurements and further insights
on the origin and the role of turbulence in clusters, groups of galaxies,
and elliptical galaxies.

The current techniques are partly limited by systematics associated with the 
spatial line broadening,
but the future ASTRO-H and ATHENA missions 
will dramatically improve the method due to their high broadband spectral resolution
and to the absence of issues concerning the spurious spatial broadening.
They will also enable spatially-resolved turbulence measurements.

\begin{acknowledgements}
This work is based on observations obtained with XMM-\textit{Newton}, an
ESA science mission funded by ESA Member States and the USA (NASA).
YYZ acknowledges the BMWi DLR grant 50~OR~1304.{We thank Electra Panagoulia
for kindly providing the values of central entropy
and the anonymous referee for very useful comments on the paper.}
\end{acknowledgements}

\begin{figure*}
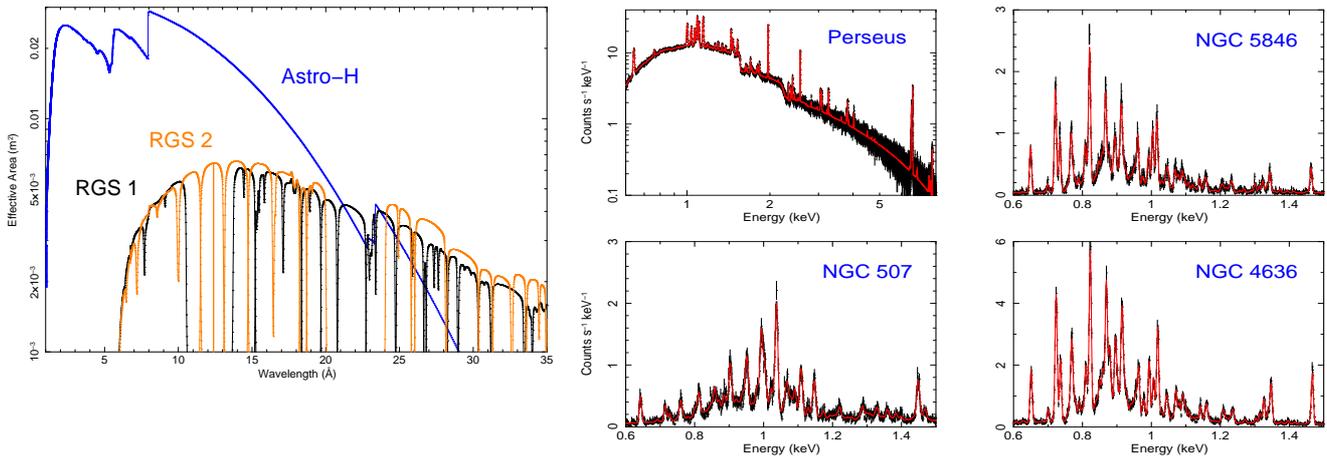

      \subfigure{ 
      \includegraphics[bb=118 78 533 689, width=5cm, angle=-90]{RGS_AstroH_effarea.ps}}
      \subfigure{ 
      \includegraphics[bb=65 45 555 755, height=10.cm,width=6cm, angle=-90]{4clusters_AstroH_sim_Vedit.ps}}
  \vspace{-0.3cm}
      \caption{\textit{Left panel}: RGS 1 and 2 first order and ASTRO-H SXS effective area.
               \textit{Right panel}: ASTRO-H SXS 100\,ks simulations for the bright Perseus cluster and three groups of galaxies
               zoomed in the RGS energy band (see Sect.\,\ref{sec:simulations}). 
               We adopted the 3.4' RGS models as a template (see Fig.\,\ref{fig:rgs_fits3}).}
          \label{fig:simulations}
\end{figure*}
  

\bibliographystyle{aa}
\bibliography{bibliografia} 

  \vspace{-0.3cm}

\appendix

\section{Maps, spectra, and velocity limits}
\label{sec:appendix}

We have placed the MOS\,1 images, the RGS spectra, and the velocity
limits in this section to unburden the paper reading.

The MOS\,1 images (Fig.\,\ref{fig:mos1}) are obtained by stacking the Fe-L (10--14\,{\AA}) band images 
extracted in each exposure (see Sect.\,\ref{sec:spatial_profile}).
We zoomed on the central 10' region.

In Table\,\ref{table:velocity_results}, we quote all our velocity results.
In Fig.\,\ref{fig:velocities_comparison} (\textit{\textit{left panel}}), we compare the 
$2\,\sigma$ upper limits on the velocity broadening 
as measured for the 3.4' and 0.8' regions by subtracting the MOS\,1 spatial profiles.
In Fig.\,\ref{fig:velocities_comparison} (\textit{right panel}), we compare the 0.8' $2\,\sigma$ upper limits
estimated by subtracting the adopted and the best-fit, spatial-line-broadening
(see Sect.\,\ref{sec:results_combined}). 
In Table\,\ref{table:velocity_results}, we also show the 
total line widths as result of spatial plus Doppler broadening
with their 68\% uncertainties.
These total widths are clearly dominated by the spatial broadening.

{In Table\,\ref{table:physical_properties}, we report the values of $r_{500}$ and $K_0$ 
adopted with with the physical scales, the RGS temperatures estimated with an isothermal model,
the conservative upper limits on the velocities and the Mach numbers, the limits
separately measured for the \ion{O}{viii}, \ion{Fe}{xvii}, and \ion{Fe}{xx-to-xxiv} emission lines,
and finally the velocity limits for the two CIE components where it was possible to fit 
them separately. The separate \ion{Fe}{xvii} and 2-T fits were accessible only for a very limited sample
of sources with both strong high- and low-ionization Fe lines (see Sect.\,\ref{sec:temperature}).
In Fig.\,\ref{fig:turb_vs_temp_new}, 
we compare the velocity upper limits estimated with both the standard and the conservative methods
with the average RGS temperature measured with an isothermal model, the sound speed, and 
the fractions of thermal energy stored in turbulence. 
In Fig.\,\ref{fig:turb_vs_temp_scaled} (\textit{left panel}), we show the Mach number
as a function of the temperature with the pointsize and, it is color coded according
to the physical scale and the central entropy, respectively.
The lines show the average Mach number calculated within particular ranges of physical scales.
In Fig.\,\ref{fig:turb_vs_temp_scaled} (\textit{left panel}), we show the Mach number
scaled by the 1/3rd power of the physical scale, assuming Kolmogorov turbulence
(see also Sect.\,\ref{sec:turbulence}).}

In Figs.\,\ref{fig:rgs_fits}, \ref{fig:rgs_fits2}, and \ref{fig:rgs_fits3},
we show the RGS spectra extracted in the 3.4' 
cross-dispersion region (see also Sect.\,\ref{sec:rgs_regions}). 
The spectra were combined with the \textit{rgscombine} task 
for plotting purposes.
We have adapted the spectral model from the bestfit obtained for
the parallel modeling of the individual exposures. 
Most spectra were fitted with a multi-temperature, two-\textit{cie} model 
(see Sect.\,\ref{sec:spectral_modeling}). A few of them require just a single \textit{cie} component.
The spectral modeling involved the $7-28$\,{\AA} band, but  
we focus on the shorter $10-21$\,{\AA} band containing the Fe, Ne, and O lines.

 \begin{figure*}
  \begin{center}
      \subfigure{ 
      \includegraphics[bb=15 15 590 445, width=18.cm]{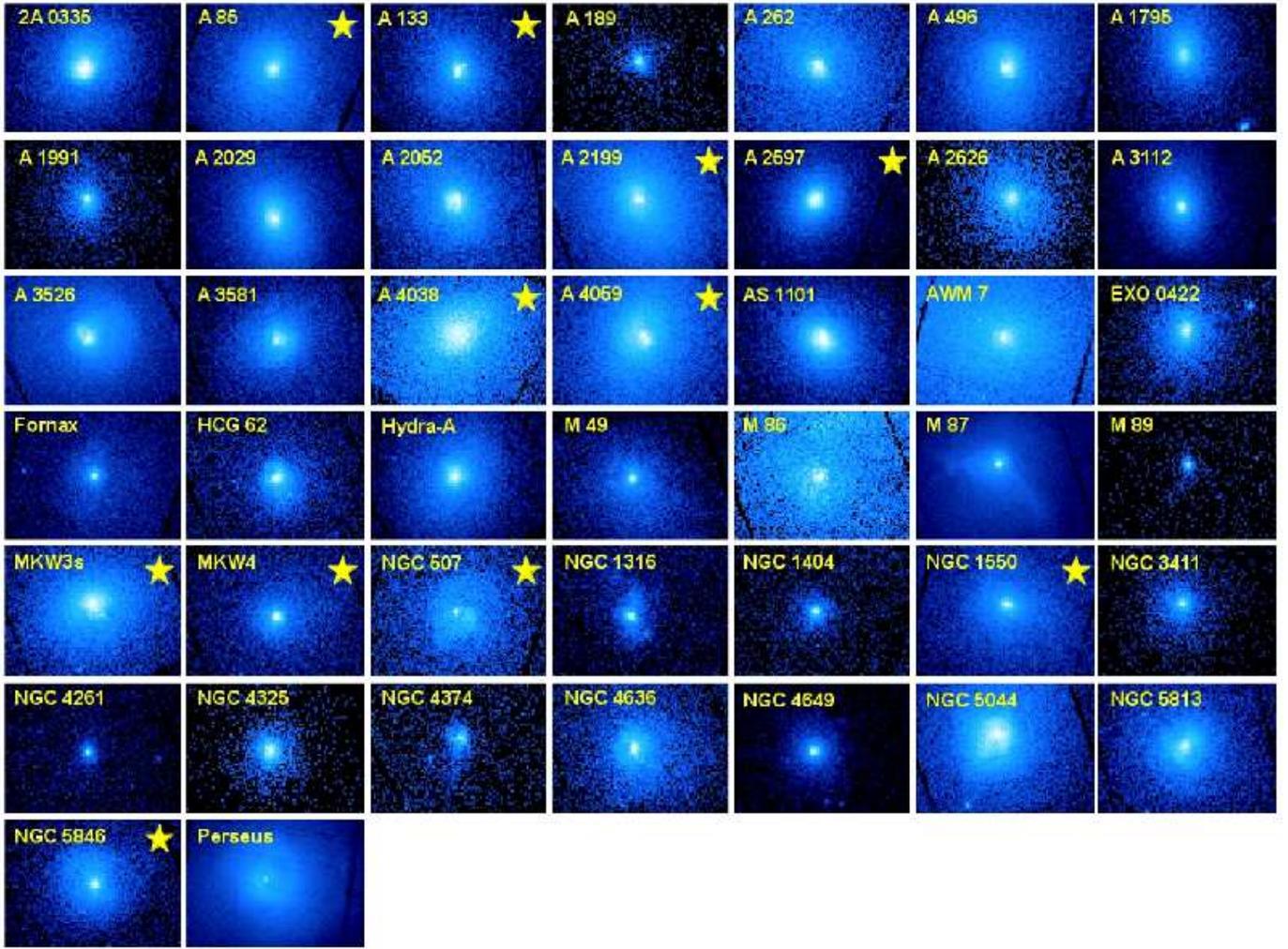}}
      \vspace{-0.2cm}
      \caption{MOS 1 stacked Fe-L band images: Central 10' $\times$ 10' region. 
               {The starred clusters are part of our new campaign}
               (see also Sect.~\ref{sec:spatial_profile}).}
          \label{fig:mos1}
  \end{center}
      \vspace{-0.3cm}
\end{figure*}

\begin{figure*}
      \subfigure{ 
      \includegraphics[bb=65 85 530 730, height=9.cm,width=7.5cm, angle=90]{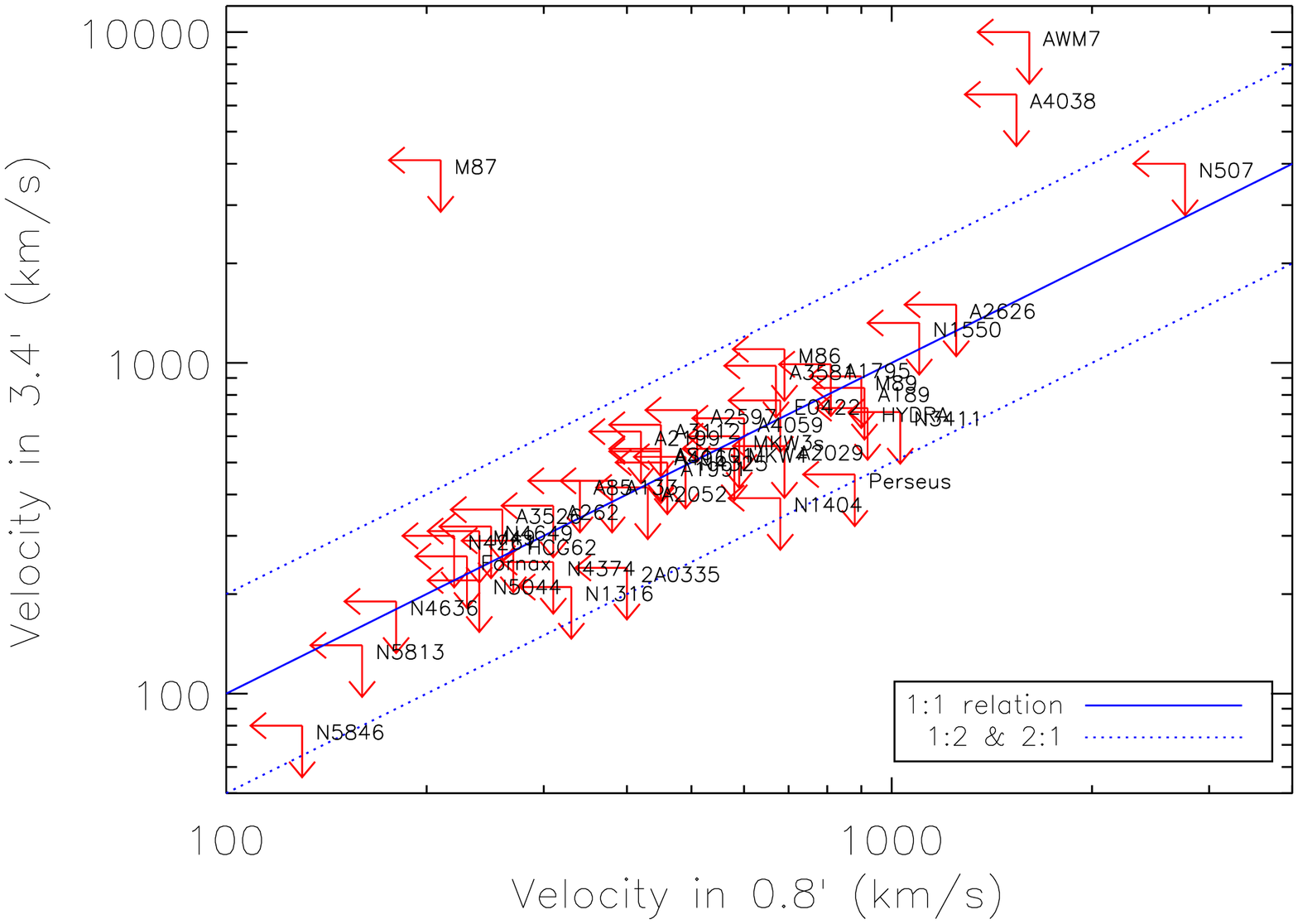}}
      \subfigure{ 
      \includegraphics[bb=65 85 530 730, height=9.cm,width=7.5cm, angle=90]{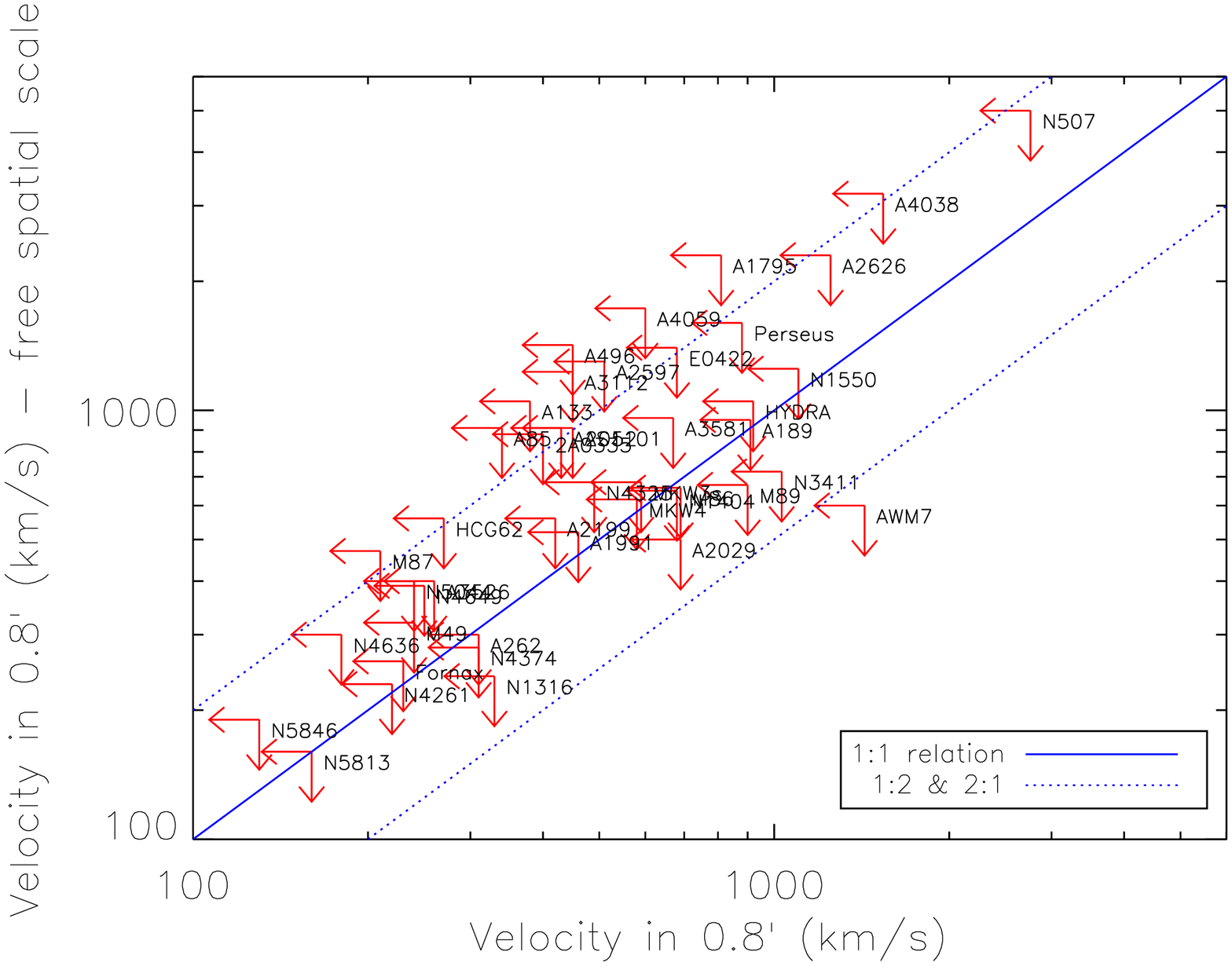}}
      \vspace{-0.25cm}
      \caption{\textit{Left panel}: Velocity broadening at $2\sigma$ upper limits for the (-1.7',+1.7') 
               and the (-0.4',+0.4') regions at comparison. The spatial broadening was removed through 
               the MOS\,1 surface brightness profiles. 
               \textit{Right panel}: (-0.4',+0.4') velocity $2\sigma$ 
               upper limits compared with those estimated in the same region
               but with the variable best-fit, spatial broadening  
               (scale parameter, \textit{s}, is free in the \textit{lpro} component, 
               see Sect.\,\ref{sec:results_combined}
               and Table\,\ref{table:velocity_results}).}
          \label{fig:velocities_comparison}
\end{figure*}

\begin{figure*}
  \begin{center}
      \subfigure{ 
      \includegraphics[bb=65 85 530 730, width=10cm, angle=90]{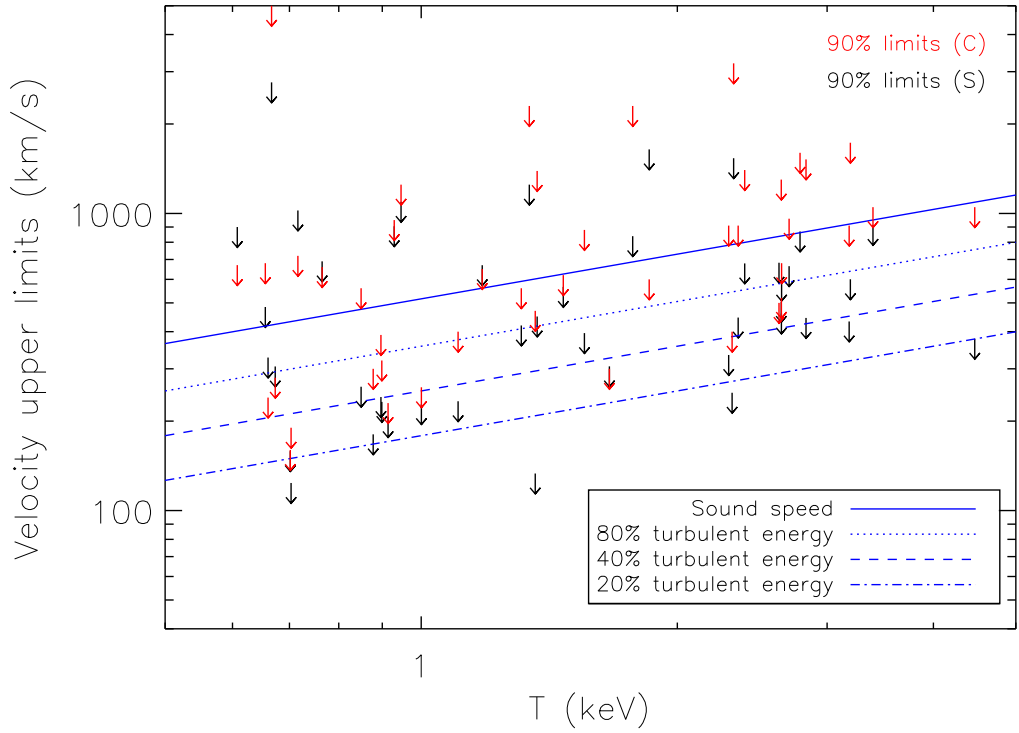}}
      \caption{90\% upper limits on velocity broadening obtained in the 0.8' region 
               versus RGS temperature (the red arrows provide the conservative limits measured 
               with the best-fit spatial broadening, see Sect.\,\ref{sec:results_combined}).
               The sound speed and the fractions of thermal energy in turbulence are shown.
               }
          \label{fig:turb_vs_temp_new}
  \end{center}
\end{figure*}

\begin{figure*}
      \subfigure{ 
      \includegraphics[bb=65 85 530 730, height=9.cm,width=7.5cm, angle=90]{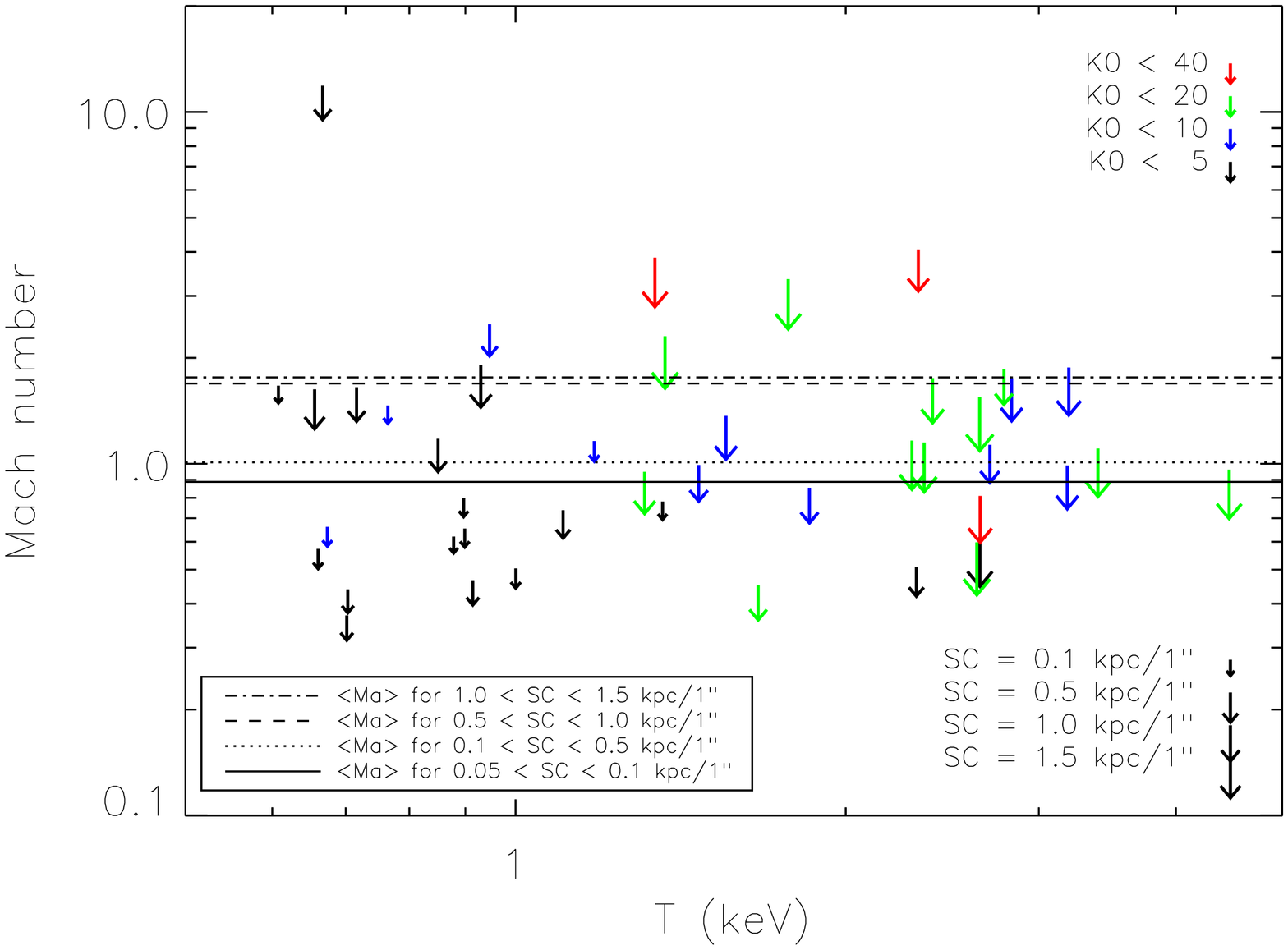}}
      \subfigure{ 
      \includegraphics[bb=65 85 530 730, height=9.cm,width=7.5cm, angle=90]{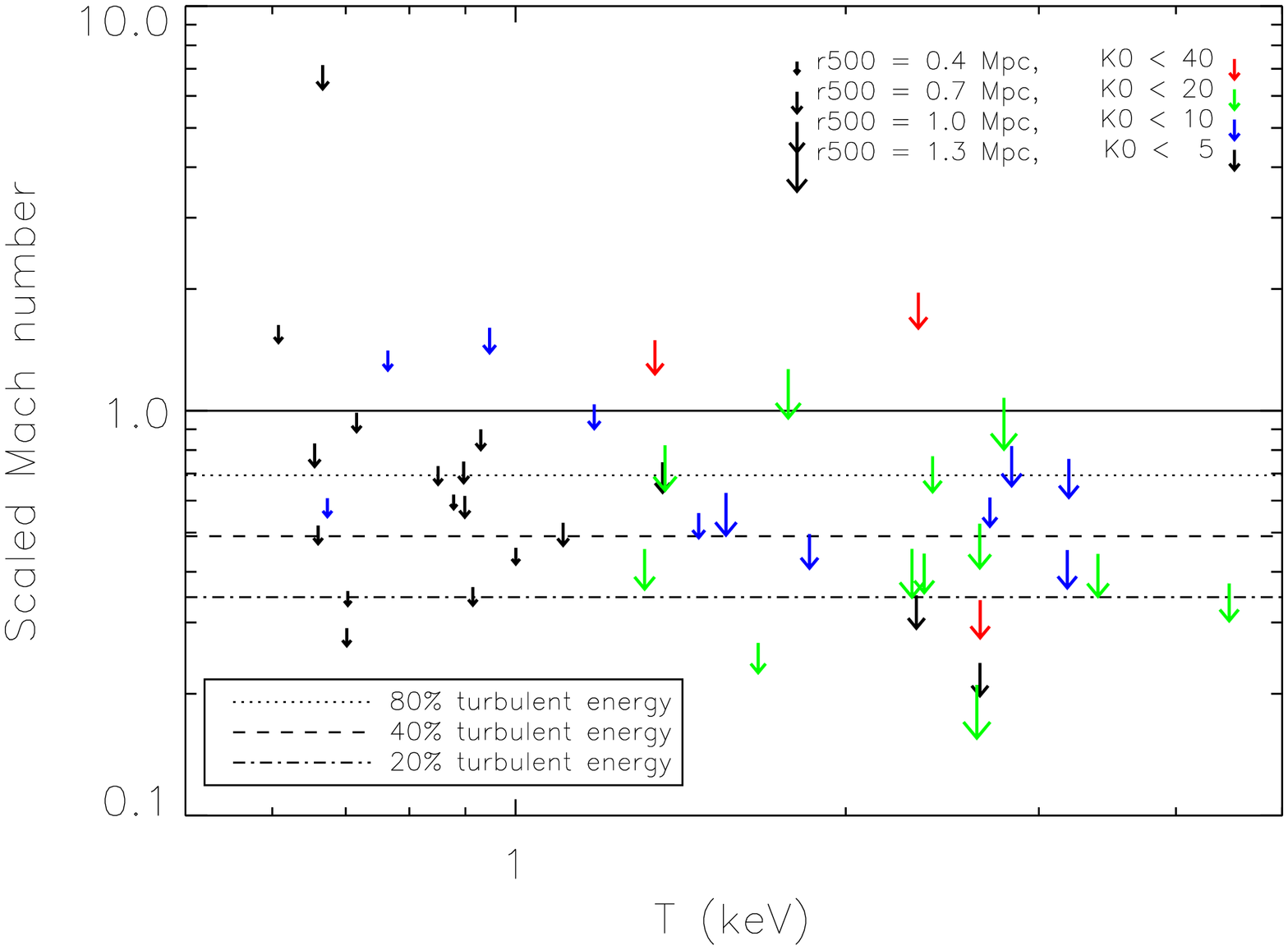}}
  \vspace{-0.4cm}
      \caption{{\textit{Left panel}: 90\% conservative upper limits on Mach number versus temperature.
               Point size refers to the physical scale and the color is coded according to the 
               central entropy, $K_0$, in units of keV cm$^{-2}$. 
               The lines show the upper limits on Mach number averaged in the four ranges
               of physical scales to underline its dependence on the source distance.
               \textit{Right panel}: Same as before, but here the limits on Mach number are scaled by the 1/3rd power 
               of the physical scale for a comparison between sources at different redshift 
               (assuming Kolmogorov turbulence, see Sect.\,\ref{sec:turbulence}).}}
          \label{fig:turb_vs_temp_scaled}
\end{figure*}




\begin{table*}
\caption{Velocity broadening upper limits and total line widths.}  
\label{table:velocity_results}      
\renewcommand{\arraystretch}{1.3}
 \small\addtolength{\tabcolsep}{+2pt}
 
\scalebox{1}{%
\begin{tabular}{c| c c c| c c c| c c}     
                                     & \multicolumn{3}{c|}{3.4' RGS extraction region}    & \multicolumn{3}{c}{0.8' RGS extraction region}     & \multicolumn{2}{c}{0.8' free spatial broadening}    \\  
Source                               &  $1\sigma$\,$^{(a)}$     &  $2\sigma$\,$^{(a)}$     & FWHM\,$^{(c)}$  & $1\sigma$\,$^{(a)}$      & $2\sigma$\,$^{(a)}$      &  FWHM\,$^{(c)}$  &  $1\sigma$\,$^{(b)}$     &  $2\sigma$\,$^{(b)}$     \\  
\hline                                                                 
\multirow{1}{*}{{2A0335+096}} &   80           &   240          & 2920 $\pm$ 140  &    150         &   400          &  2240 $\pm$ 140  &  570           &  880           \\
\multirow{1}{*}{{A 85}}       &  290           &   440          & 2100 $\pm$ 500  &    220         &   340          &  720  $\pm$ 160  &  580           &  910           \\
\multirow{1}{*}{{A 133}}      &  280           &   440          & 1610 $\pm$ 220  &    230         &   380          &  1170 $\pm$ 140  &  980           &  1050          \\
\multirow{1}{*}{{A 189}}      &  530           &   840          & 1980 $\pm$ 320  &    700         &   910          &  1350 $\pm$ 190  &  600           &  950           \\
\multirow{1}{*}{{A 262}}      &  180           &   370          & 4570 $\pm$ 410  &    180         &   310          &  2340 $\pm$ 450  &  170           &  300           \\
\multirow{1}{*}{{A 496}}      &  310           &   540          & 3150 $\pm$ 220  &    270         &   450          &  1970 $\pm$ 170  &  1250          &  1520          \\
\multirow{1}{*}{{A 1795}}     &  610           &   990          & 2590 $\pm$ 320  &    540         &   810          &  1687 $\pm$ 320  &  2100          & 2300           \\
\multirow{1}{*}{{A 1991}}     &  270           &   500          &  830 $\pm$ 250  &    250         &   460          &   480 $\pm$ 230  &  300           &  520           \\
\multirow{1}{*}{{A 2029}}     &  330           &   560          & 1470 $\pm$ 700  &    410         &   690          &   720 $\pm$ 260  &  330           &  500           \\
\multirow{1}{*}{{A 2052}}     &  190           &   420          & 2800 $\pm$ 500  &    180         &   430          &  2289 $\pm$ 220  &  570           &  910           \\
\multirow{1}{*}{{A 2199}}     &  380           &   620          & 3800 $\pm$ 300  &    240         &   420          &  2540 $\pm$ 230  &  330           &  560           \\
\multirow{1}{*}{{A 2597}}     &  480           &   720          & 1730 $\pm$ 270  &    300         &   510          &  1130 $\pm$ 170  & 1200           & 1300           \\
\multirow{1}{*}{{A 2626}}     &  900           &  1500          & 3300 $\pm$ 900  &    850         &  1250          &  2440 $\pm$ 610  & 1600           & 2300           \\
\multirow{1}{*}{{A 3112}}     &  420           &   650          & 2000 $\pm$ 170  &    270         &   450          &  1120 $\pm$ 130  & 1240           & 1390           \\
\multirow{1}{*}{{A 3526}}     &  260           &   360          & 3930 $\pm$ 110  &    190         &   260          &  2470 $\pm$ 160  &  290           &  400           \\
\multirow{1}{*}{{A 3581}}     &  610           &   980          & 4390 $\pm$ 220  &    410         &   670          &  2980 $\pm$ 260  &  700           &  960           \\
\multirow{1}{*}{{A 4038}}     & 5180           &  6480          & 8100 $\pm$ 800  &    930         &  1540          &  4700 $\pm$ 750  & 2600           & 3200           \\
\multirow{1}{*}{{A 4059}}     &  420           &   680          & 2620 $\pm$ 250  &    360         &   600          &  1860 $\pm$ 180  & 1490           & 1730           \\
\multirow{1}{*}{{AS 1101}}    &  340           &   550          & 2650 $\pm$ 200  &    270         &   450          &  1790 $\pm$ 180  &  600           &  910           \\
\multirow{1}{*}{{AWM 7}}      & 8000           & 10000          &12600 $\pm$ 700  &   1000         &  1430          & 10500 $\pm$ 700  &  150           &  600           \\
\multirow{1}{*}{{EXO 0422}}   &  480           &   770          & 1520 $\pm$ 280  &    400         &   680          &  1020 $\pm$ 230  & 1300           & 1400           \\
\multirow{1}{*}{{Fornax}}     &  160           &   260          & 2500 $\pm$ 500  &    160         &   230          &   820 $\pm$ 100  &  170           &  260           \\
\multirow{1}{*}{{HCG 62}}     &  150           &   290          & 2180 $\pm$ 190  &    140         &   270          &  1420 $\pm$  90  &  350           &  560           \\
\multirow{1}{*}{{Hydra-A}}    &  360           &   730          & 3990 $\pm$ 400  &    550         &   920          &  2420 $\pm$ 370  &  460           & 1050           \\
\multirow{1}{*}{{M 49}}       &  190           &   310          & 3180 $\pm$ 190  &    150         &   240          &  1350 $\pm$ 100  &  220           &  320           \\
\multirow{1}{*}{{M 86}}       &  650           &  1100          & 7500 $\pm$ 400  &    400         &   690          &  5960 $\pm$ 490  &  390           &  660           \\
\multirow{1}{*}{{M 87}}       & 3800           &  4100          & 5290 $\pm$ 150  &    100         &   210          &  3850 $\pm$ 150  &  170           &  470           \\
\multirow{1}{*}{{M 89}}       &  740           &   910          & 1700 $\pm$ 180  &    710         &   900          &  1300 $\pm$ 190  &  460           &  670           \\
\multirow{1}{*}{{MKW 3s}}     &  350           &   600          & 4010 $\pm$ 570  &    360         &   590          &  2170 $\pm$ 250  &  430           &  680           \\
\multirow{1}{*}{{MKW 4}}      &  310           &   550          & 3620 $\pm$ 330  &    360         &   580          &  2040 $\pm$ 140  &  380           &  620           \\
\multirow{1}{*}{{NGC 507}}    & 2500 $\pm$ 1000& 2500 $\pm$ 1500& 6100 $\pm$ 900  &1500 $\pm\;_{1300}^{\;800}$& 2800    &  4600 $\pm$ 500  & 3000 $\pm$ 1200& 5000           \\
\multirow{1}{*}{{NGC 1316}}   &   20           &   210          & 1770 $\pm$ 120  &    110         &   330          &  1160 $\pm$  90  &   90           &  240           \\
\multirow{1}{*}{{NGC 1404}}   &  250           &   390          & 1930 $\pm$ 130  &    520         &   680          &  1470 $\pm$ 130  &  540           &  650           \\
\multirow{1}{*}{{NGC 1550}}   &  860           &  1320          & 7500 $\pm$ 700  &    890         &  1100          &  3750 $\pm$ 250  & 1130           & 1250           \\
\multirow{1}{*}{{NGC 3411}}   &  290           &   710          & 3830 $\pm$ 360  &    610         &  1030          &  2230 $\pm$ 310  &  370           &  720           \\
\multirow{1}{*}{{NGC 4261}}   &  160           &   300          &  760 $\pm$ 120  &    140         &   220          &   470 $\pm$ 350  &  160           &  230           \\
\multirow{1}{*}{{NGC 4325}}   &  300           &   520          & 1570 $\pm$ 240  &    310         &   490          &  1060 $\pm$ 180  &  440           &  680           \\
\multirow{1}{*}{{NGC 4374}}   &  120           &   250          & 1770 $\pm$ 140  &    180         &   310          &  1400 $\pm$ 490  &  160           &  280           \\
\multirow{1}{*}{{NGC 4636}}   &  140           &   190          & 2410 $\pm$  70  &    130         &   180          &  1660 $\pm$  60  &  190           &  300           \\
\multirow{1}{*}{{NGC 4649}}   &  210           &   320          & 1520 $\pm$  80  &    130         &   250          &  1050 $\pm$  70  &  280           &  390           \\
\multirow{1}{*}{{NGC 5044}}   &   80           &   220          & 4170 $\pm$ 180  &    100         &   240          &  3100 $\pm$ 200  &  140           &  400           \\
\multirow{1}{*}{{NGC 5813}}   &  100           &   140          & 4080 $\pm$  80  &    120         &   160          &  2980 $\pm$ 120  &  110           &  160           \\
\multirow{1}{*}{{NGC 5846}}   &   40           &    80          & 2100 $\pm$  60  &     20         &   130          &  1670 $\pm$  50  &  100           &  190           \\
\multirow{1}{*}{{Perseus}}    &  320           &   460          & 4950 $\pm$ 140  &    620         &   880          &  4381 $\pm$ 390  & 1000           & 1600           \\
\hline                                                                                                                                                            
\end{tabular}}
      \vspace{0.1cm}
 \\
$^{(a)}$ Velocity $1\sigma$ and $2\sigma$ upper limits for the RGS spectra extracted within the 3.4' and 0.8' regions 
centered on the source emission peak. \\ For NGC\,507, we quote the best-fit velocities.
The spatial line broadening was subtracted through the MOS\,1 maps 
(see Figs.\,\ref{fig:turbulence2} and \ref{fig:velocities_comparison} \textit{\textit{left panel}}).\\
$^{(b)}$ Velocity limits for the 0.8' region obtained by subtracting the best-fit spatial line broadening 
(see Sect.\,\ref{sec:results_combined} and Fig.\,\ref{fig:velocities_comparison} \textit{right panel}).\\
$^{(c)}$ Total (spatial\,+\,velocity) line widths with $1\sigma$ uncertainties (see Sect.\,\ref{sec:tests}).\\
      \vspace{-0.5cm}
\end{table*}
      \vspace{-0.2cm}

\newpage
\clearpage

\begin{table*}
\caption{Additional results and physical properties.}  
      \vspace{-0.25cm}
\label{table:physical_properties}      
\renewcommand{\arraystretch}{1.25}
 \small\addtolength{\tabcolsep}{-2pt}
 
\scalebox{1}{%
\begin{tabular}{c| c c c c c c c| c c c c| c c c| c c}     
 & \multicolumn{7}{c|}{Physical properties\,$^{(a)}$}  & \multicolumn{4}{c|}{RGS-band $2\sigma$ limits\,$^{(b)}$} & \multicolumn{3}{c|}{Emission-line $2\sigma$ limits\,$^{(c)}$} & \multicolumn{2}{c}{2-T $2\sigma$ limits\,$^{(d)}$}  \\  
Source                        & d(Mpc)& kpc/1"& $r_{500}$   &  $K_0$    & T     & $c_S$  & $Ma_{REQ}$ & $v$   & $v_{SC}$ & $Ma$ & $Ma_{SC}$ & \ion{O}{viii} & \ion{Fe}{xvii} & \ion{Fe}{xx+} & $v_1$ & $v_2$ \\  
\hline                                                                                                                                                                                   
\multirow{1}{*}{{2A0335+096}} & 149   & 0.72  & 1.05   &  7.1   & 1.6   &   640  & 0.27   &  880  &    400   &  1.37 & 0.63  &    580  &   --    &    530   &    --    &   --    \\
\multirow{1}{*}{{A 85}}       & 238   & 1.15  & 1.21   & 12.5   & 2.3   &   780  & 0.28   &  910  &    360   &  1.16 & 0.46  &    480  &   --    &    570   &    --    &   --    \\
\multirow{1}{*}{{A 133}}      & 243   & 1.18  & 0.94   & 17.3   & 4.5   &  1090  & 0.13   & 1050  &    410   &  0.96 & 0.37  &    460  &   --    &    603   &    --    &   --    \\
\multirow{1}{*}{{A 189}}      & 137   & 0.66  & 0.50   &  4.0   & 0.9   &   500  & 0.18   &  950  &    450   &  1.91 & 0.90  &    650  &   --    &   3010   &    --    &   --    \\
\multirow{1}{*}{{A 262}}      &  69   & 0.34  & 0.74   & 10.6   & 1.7   &   670  & 0.17   &  300  &    180   &  0.45 & 0.27  &    510  &   --    &    790   &    --    &   --    \\
\multirow{1}{*}{{A 496}}      & 140   & 0.68  & 1.00   &  8.9   & 2.8   &   870  & 0.18   & 1520  &    710   &  1.75 & 0.82  &    480  &   --    &    710   &    --    &   --    \\
\multirow{1}{*}{{A 1795}}     & 264   & 1.28  & 1.22   & 19.0   & 1.8   &   690  & 0.36   & 2300  &    870   &  3.35 & 1.27  &    930  &   --    &   1750   &    --    &   --    \\
\multirow{1}{*}{{A 1991}}     & 251   & 1.22  & 0.82   &  1.5   & 2.7   &   840  & 0.21   &  520  &    200   &  0.62 & 0.24  &    630  &   --    &    900   &    --    &   --    \\
\multirow{1}{*}{{A 2029}}     & 328   & 1.59  & 1.33   & 10.5   & 2.6   &   840  & 0.23   &  500  &    180   &  0.60 & 0.21  &    810  &   --    &   2310   &    --    &   --    \\
\multirow{1}{*}{{A 2052}}     & 149   & 0.72  & 0.95   &  9.5   & 3.2   &   920  & 0.21   &  910  &    420   &  0.99 & 0.45  &    600  &   --    &    690   &    --    &   --    \\
\multirow{1}{*}{{A 2199}}     & 129   & 0.63  & 1.00   & 13.3   & 1.3   &   590  & 0.28   &  560  &    270   &  0.95 & 0.46  &    540  &   --    &   1010   &    --    &   --    \\
\multirow{1}{*}{{A 2597}}     & 365   & 1.77  & 1.11   & 10.6   & 2.6   &   840  & 0.23   & 1300  &    440   &  1.55 & 0.53  &    500  &   --    &    990   &    --    &   --    \\
\multirow{1}{*}{{A 2626}}     & 245   & 1.19  & 0.84   & 23.2   & 1.3   &   600  & 0.23   & 2300  &    890   &  3.85 & 1.49  &    --   &   --    &    --    &    --    &   --    \\
\multirow{1}{*}{{A 3112}}     & 315   & 1.53  & 1.13   & 11.4   & 1.4   &   600  & 0.25   & 1390  &    500   &  2.30 & 0.82  &    500  &   --    &    570   &    --    &   --    \\
\multirow{1}{*}{{A 3526}}     &  44   & 0.21  & 0.83   &  2.3   & 2.3   &   790  & 0.20   &  400  &    280   &  0.51 & 0.35  &    390  &   230   &    730   &    530   &    240  \\
\multirow{1}{*}{{A 3581}}     &  91   & 0.44  & 0.72   &  9.5   & 2.7   &   850  & 0.14   &  960  &    520   &  1.13 & 0.61  &   1150  &   --    &   1060   &    --    &    --   \\
\multirow{1}{*}{{A 4038}}     & 128   & 0.62  & 0.89   & 37.9   & 2.3   &   790  & 0.18   & 3200  &   1540   &  4.06 & 1.96  &   1540  &   --    &   5290   &    --    &    --   \\
\multirow{1}{*}{{A 4059}}     & 215   & 1.04  & 0.96   &  7.1   & 3.2   &   920  & 0.19   & 1730  &    700   &  1.88 & 0.76  &   1090  &   --    &   1120   &    --    &    --   \\
\multirow{1}{*}{{AS 1101}}    & 248   & 1.20  & 0.98   & 10.4   & 2.4   &   800  & 0.23   &  910  &    350   &  1.15 & 0.44  &    530  &   --    &    690   &    --    &    --   \\
\multirow{1}{*}{{AWM 7}}      &  73   & 0.35  & 0.86   &  8.4   & 1.9   &   700  & 0.22   &  600  &    350   &  0.85 & 0.50  &   1650  &   --    &    --    &    --    &    --   \\
\multirow{1}{*}{{EXO 0422}}   & 167   & 0.81  & 0.89   & 13.8   & 2.4   &   800  & 0.18   & 1400  &    620   &  1.75 & 0.77  &   1140  &   --    &   1170   &    --    &    --   \\
\multirow{1}{*}{{Fornax}}     &  19   & 0.09  & 0.40   &  2.6   & 1.0   &   520  & 0.16   &  260  &    240   &  0.50 & 0.46  &    390  &   300   &    520   &    --    &    --   \\
\multirow{1}{*}{{HCG 62}}     &  60   & 0.29  & 0.46   &  3.4   & 0.8   &   480  & 0.23   &  560  &    350   &  1.18 & 0.73  &    910  &   320   &   1350   &   2370   &    260  \\
\multirow{1}{*}{{Hydra-A}}    & 222   & 1.08  & 1.07   & 13.3   & 3.4   &   950  & 0.21   & 1050  &    420   &  1.10 & 0.44  &    700  &   --    &   1050   &    --    &    --   \\
\multirow{1}{*}{{M 49}}       &  17.1 & 0.08  & 0.53   &  0.9   & 0.9   &   490  & 0.17   &  320  &    300   &  0.65 & 0.62  &    520  &   340   &    940   &   1050   &    240  \\
\multirow{1}{*}{{M 86}}       &  16   & 0.08  & 0.49   &  8.0   & 0.8   &   450  & 0.22   &  660  &    640   &  1.46 & 1.41  &    --   &   --    &    --    &    --    &    --   \\
\multirow{1}{*}{{M 87}}       &  16.4 & 0.08  & 0.75   &  3.5   & 1.4   &   600  & 0.25   &  470  &    450   &  0.78 & 0.75  &    400  &   --    &    430   &    --    &    --   \\
\multirow{1}{*}{{M 89}}       &  15.3 & 0.07  & 0.44   &  3.0   & 0.6   &   400  & 0.20   &  670  &    660   &  1.67 & 1.63  &   1330  &   960   &   1430   &    --    &    --   \\
\multirow{1}{*}{{MKW 3s}}     & 192   & 0.93  & 0.95   & 23.9   & 2.7   &   840  & 0.16   &  680  &    290   &  0.81 & 0.34  &    720  &   --    &   1310   &    --    &    --   \\
\multirow{1}{*}{{MKW 4}}      &  80   & 0.39  & 0.62   &  6.9   & 1.5   &   630  & 0.31   &  620  &    350   &  0.99 & 0.56  &    660  &   --    &    700   &    --    &    --   \\
\multirow{1}{*}{{NGC 507}}    &  65.5 & 0.32  & 0.60   &  1.0   & 0.7   &   420  & 0.22   & 5000  &   3010   & 11.87 & 7.15  &   4210  &  1310   &   4560   &   6540   &   1680  \\
\multirow{1}{*}{{NGC 1316}}   &  19   & 0.09  & 0.46   &  1.0   & 0.7   &   420  & 0.20   &  240  &    220   &  0.57 & 0.52  &    410  &   250   &   4780   &    --    &    --   \\
\multirow{1}{*}{{NGC 1404}}   &  20   & 0.10  & 0.61   &  5.5   & 1.2   &   560  & 0.23   &  650  &    580   &  1.16 & 1.04  &   1020  &   690   &   1180   &    --    &    --   \\
\multirow{1}{*}{{NGC 1550}}   &  53.6 & 0.26  & 0.62   &  6.6   & 0.9   &   500  & 0.22   & 1250  &    800   &  2.49 & 1.60  &    860  &   --    &   1010   &    --    &    --   \\
\multirow{1}{*}{{NGC 3411}}   &  66.3 & 0.32  & 0.47   &  3.7   & 0.7   &   440  & 0.21   &  720  &    430   &  1.65 & 0.99  &   1790  &  1520   &   1570   &    --    &    --   \\
\multirow{1}{*}{{NGC 4261}}   &  29.4 & 0.14  & 0.45   &  3.0   & 0.9   &   500  & 0.19   &  230  &    180   &  0.47 & 0.37  &    650  &   290   &    620   &    --    &    --   \\
\multirow{1}{*}{{NGC 4325}}   & 108   & 0.52  & 0.58   &  4.7   & 0.7   &   420  & 0.29   &  680  &    350   &  1.63 & 0.83  &   1690  &   540   &   1870   &    --    &    --   \\
\multirow{1}{*}{{NGC 4374}}   &  18.4 & 0.09  & 0.46   &  8.0   & 0.7   &   420  & 0.19   &  280  &    260   &  0.66 & 0.61  &    510  &   550   &    900   &    --    &    --   \\
\multirow{1}{*}{{NGC 4636}}   &  14.3 & 0.07  & 0.35   &  1.4   & 0.9   &   480  & 0.13   &  300  &    300   &  0.62 & 0.62  &    310  &   290   &   1410   &    --    &    --   \\
\multirow{1}{*}{{NGC 4649}}   &  17.3 & 0.08  & 0.53   &  4.6   & 0.9   &   490  & 0.19   &  390  &    370   &  0.80 & 0.75  &    370  &   580   &    590   &    --    &    --   \\
\multirow{1}{*}{{NGC 5044}}   &  38.9 & 0.19  & 0.56   &  2.3   & 1.1   &   540  & 0.24   &  400  &    290   &  0.74 & 0.53  &    500  &   350   &   1770   &    790   &    280  \\
\multirow{1}{*}{{NGC 5813}}   &  29.7 & 0.14  & 0.44   &  1.4   & 0.7   &   430  & 0.24   &  160  &    130   &  0.37 & 0.29  &    470  &   170   &   1690   &    --    &    --   \\
\multirow{1}{*}{{NGC 5846}}   &  26.3 & 0.13  & 0.36   &  1.8   & 0.7   &   430  & 0.39   &  190  &    160   &  0.44 & 0.36  &    320  &   190   &   1470   &    300   &    180  \\
\multirow{1}{*}{{Perseus}}    &  73.6 & 0.36  & 1.29   & 19.4   & 2.8   &   860  & 0.28   & 1600  &    926   &  1.86 & 1.08  &   1160  &   --    &   1670   &    --    &    --   \\
\hline                                                                                                                                                            
\end{tabular}}
      \vspace{0.01cm}
 \\
$^{(a)}$ $r_{500}$ (in Mpc units) and $K_0$ (in keV cm$^{-2}$ units) were taken from the ACCEPT catalog \citep{Cavagnolo2009} 
with the exceptions of M\,86 \citep{Finoguenov2004}, NGC\,1316 \citep{Tashiro2006},
NGC\,4649, and NGC\,4261 \citep{Werner2012}, A\,189, Fornax, HCG\,62, M\,49, NGC\,1550, 
NGC\,3411, NGC\,4325, NGC\,4636, NGC\,5044, NGC\,5813, and NGC\,5846 \citep{Panagoulia2014}.
NGC\,4374 and NGC\,1404 profiles were calculated following the method of \citet{Werner2012}.
The temperatures (in keV units) are estimated with the 0.8' region RGS best-fit isothermal model, 
and they are used to compute the sound speed, $c_S$, 
which is reported in units of km\,s$^{-1}$ (see Sect.\,\ref{sec:turbulence}).\\
$Ma_{REQ}$ is the Mach number which is required to make a heating--cooling balance. It is calculated with Eq.\,(\ref{eq:mach}) in Sect.\,\ref{sec:turbulence}.\\
$^{(b)}$ 90\% limits on velocity and Mach number for the 0.8' region, obtained by subtracting 
the best-fit spatial line broadening, and 90\% limits normalized by the physical scale,
assuming Kolmogorov turbulence
(see Sects.\,\ref{sec:results_combined} and \ref{sec:turbulence}; 
Figs.\,\ref{fig:turbulence2} and \ref{fig:turb_vs_temp_new}; and last column in Table\,\ref{table:velocity_results}).\\
$^{(c)}$ 90\% velocity limits separately measured for the relevant emission lines with an isothermal cluster model (see Sect.\,\ref{sec:temperature}).\\
$^{(d)}$ 90\% velocity limits separately measured for the two CIE components with the multi-temperature cluster model (see Sect.\,\ref{sec:temperature}).\\
      \vspace{-0.5cm}
\end{table*}
      \vspace{-0.2cm}

\begin{figure*}
  \begin{center}
      \subfigure{ 
      \includegraphics[bb=25 15 585 760, width=0.95\textwidth]{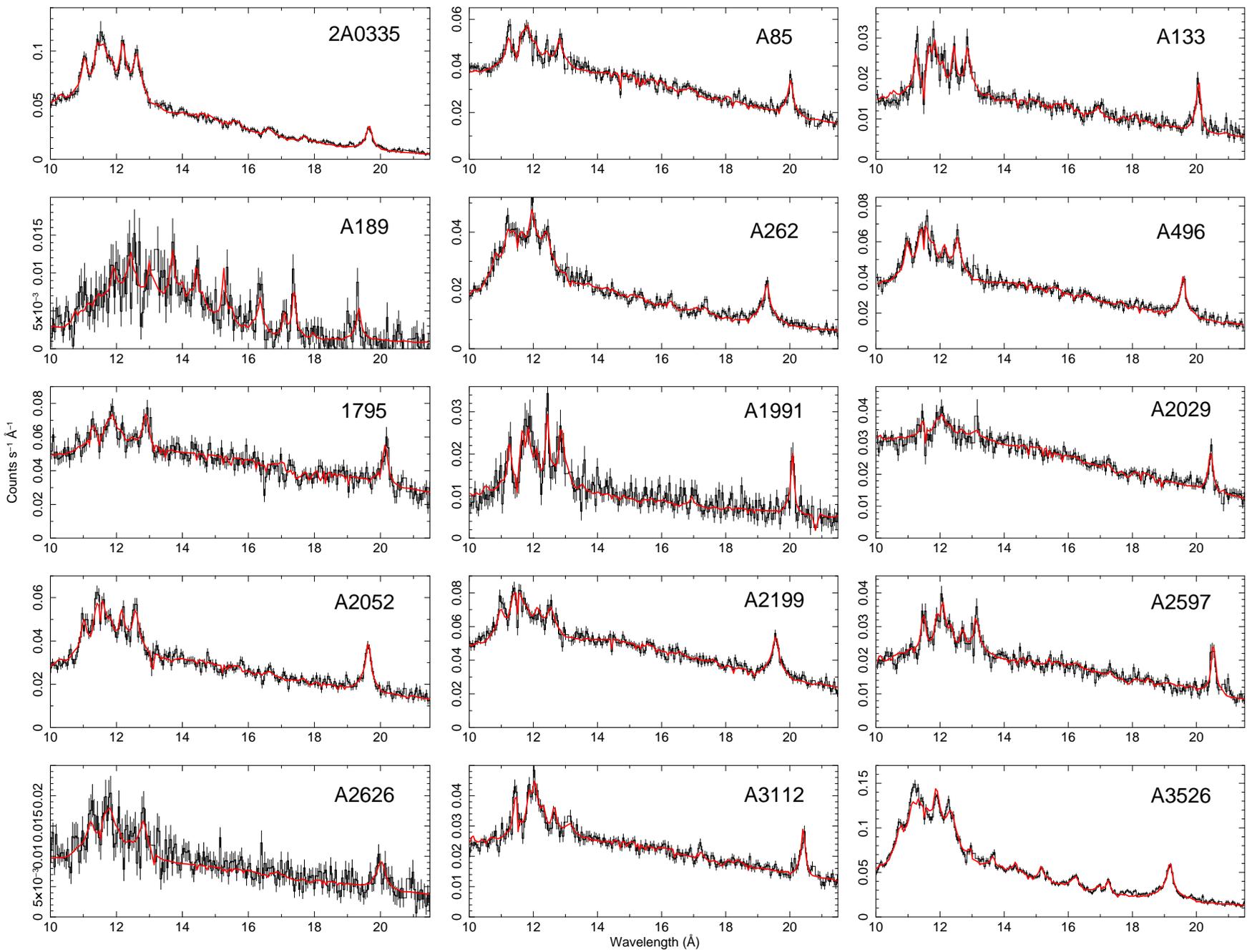}}
      \caption{RGS spectral fits for the (-1.7',+1.7') region 
               with the $7-25$\,{\AA} spatial broadening profile (Part I).
               \newline{For displaying purposes, the spectra were combined using the XMM-SAS task \textit{rgscombine}.}}
          \label{fig:rgs_fits}
  \end{center}
\end{figure*}

\begin{figure*}
  \begin{center}
      \subfigure{ 
      \includegraphics[bb=25 15 585 760, width=0.95\textwidth]{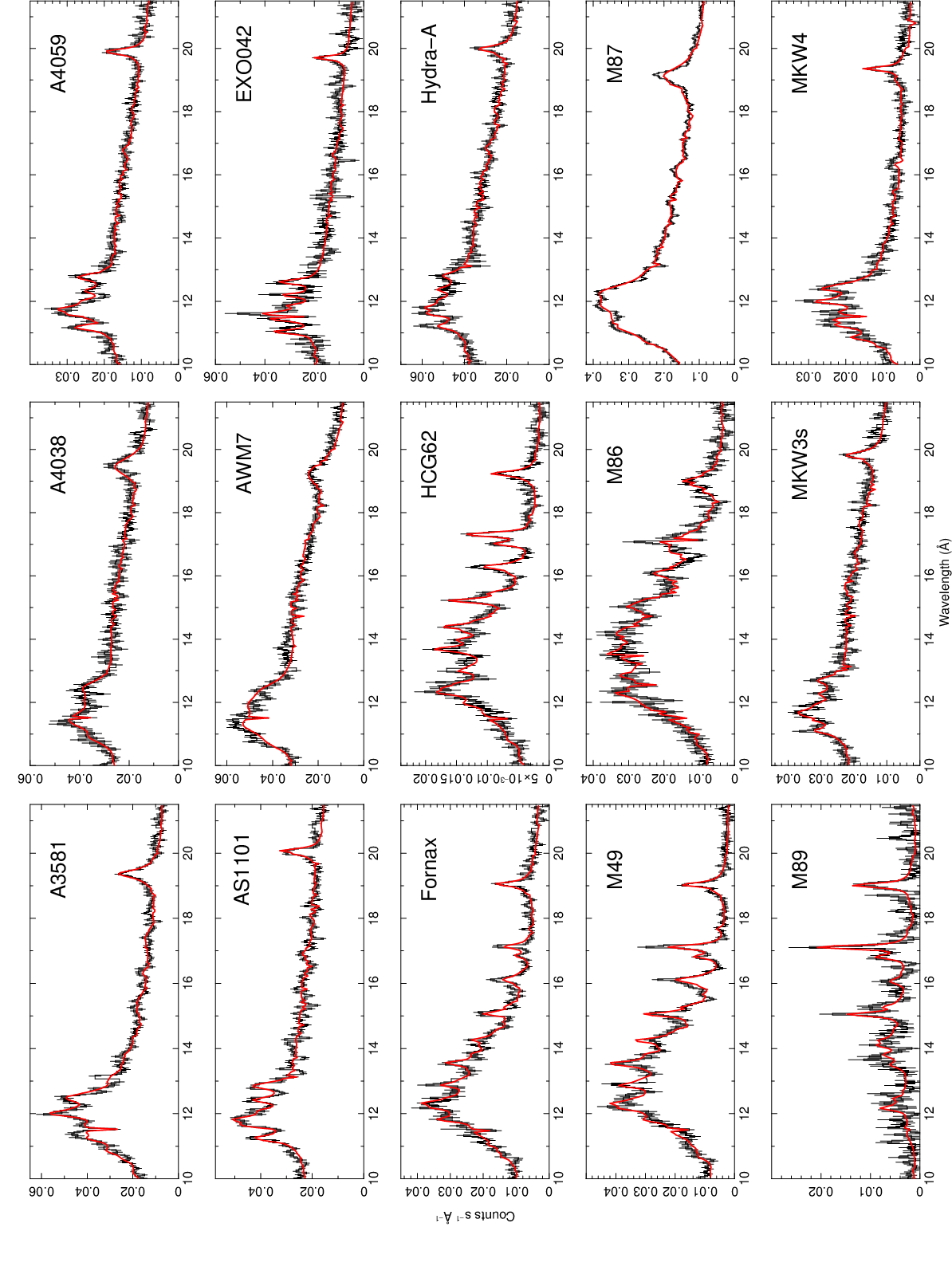}}
      \caption{RGS spectral fits for the (-1.7',+1.7') region 
               with the $7-25$\,{\AA} spatial broadening profile (Part II).
               \newline{For displaying purposes, the spectra were combined using the XMM-SAS task \textit{rgscombine}.}}
          \label{fig:rgs_fits2}
  \end{center}
\end{figure*}

\begin{figure*}
  \begin{center}
      \subfigure{ 
      \includegraphics[bb=25 15 585 760, width=0.95\textwidth]{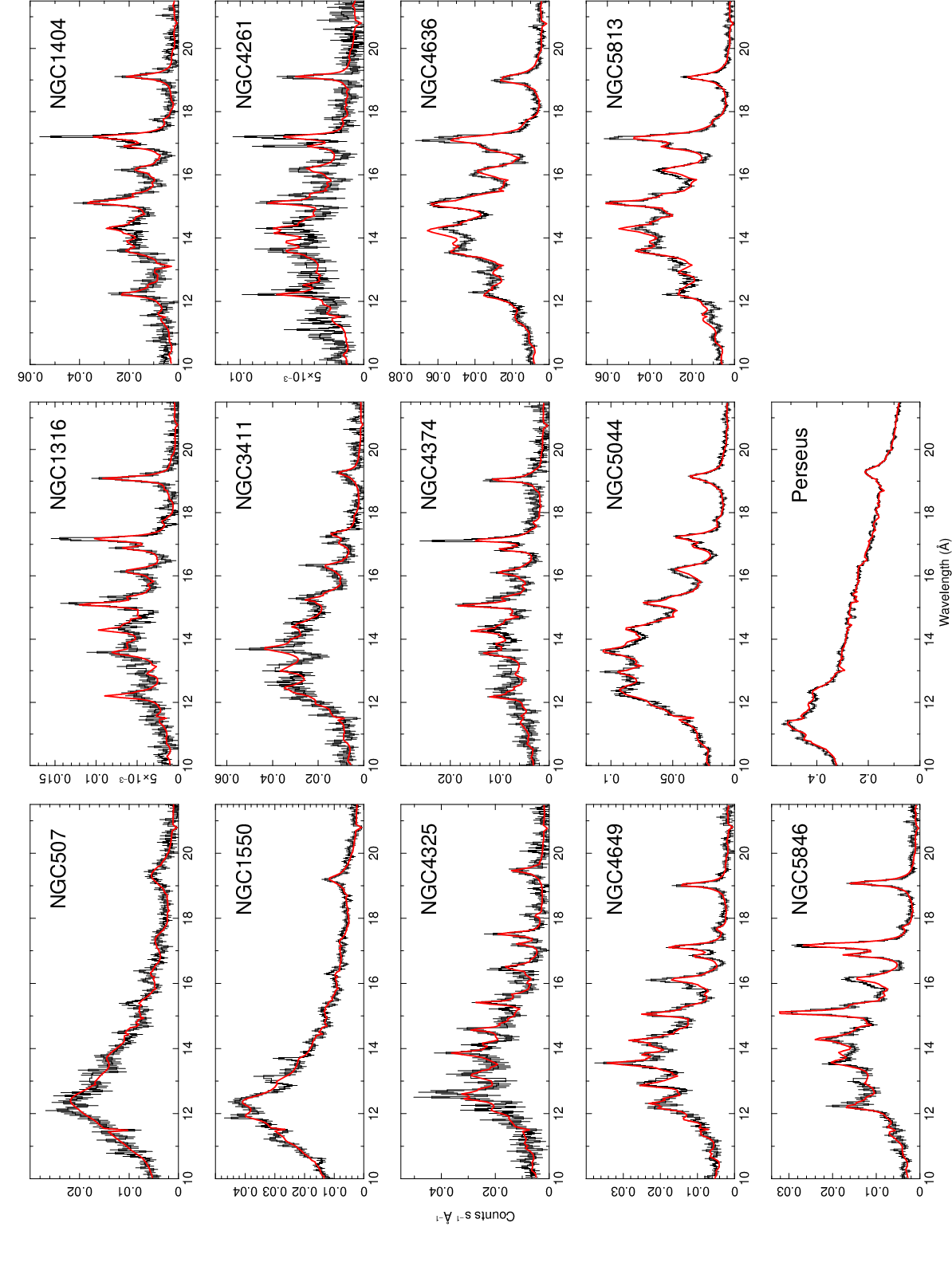}}
      \caption{RGS spectral fits for the (-1.7',+1.7') region 
               with the $7-25$\,{\AA} spatial broadening profile (Part III).
               \newline{For displaying purposes, the spectra were combined using the XMM-SAS task \textit{rgscombine}.}}
          \label{fig:rgs_fits3}
  \end{center}
\end{figure*}

\end{document}